\documentclass[aps,prb,twocolumn,floatfix,nofootinbib,superscriptaddress,longbibliography,nobibnotes]{revtex4-2}

\usepackage{amsmath}
\usepackage{amssymb}
\usepackage{mathtools}

\usepackage[T1]{fontenc}
\usepackage{amsfonts}
\usepackage{newtxtext}
\usepackage[varvw]{newtxmath}
\usepackage{dsfont}                 
\usepackage{bbold}                  
\usepackage[normalem]{ulem}         

\usepackage{array}
\usepackage{dcolumn}                

\usepackage{enumerate}
\usepackage[shortlabels]{enumitem}

\usepackage[usenames,dvipsnames,table]{xcolor}
\usepackage{graphicx}
\graphicspath{{figs/}} 

\usepackage{physics}                
\usepackage{csquotes}               
\usepackage{multirow}               
\usepackage{booktabs}               

\usepackage[caption=false]{subfig}
\usepackage[colorlinks=true,citecolor=blue,linkcolor=blue,filecolor=blue]{hyperref}
\usepackage[capitalize]{cleveref}

\usepackage{orcidlink}

\usepackage[compat=1.1.0]{tikz-feynman}

\usepackage{comment}
\usepackage[normalem]{ulem}
\usepackage{xcolor}

\definecolor{JM}{rgb}{1,0.25,0.9}

\definecolor{FM}{rgb}{0.4,0.0,1.0}

\newcommand{\sgn}{\mathop{\mathrm{sgn}}}
\renewcommand{\c}[1]{\mathcal{#1}}

\renewcommand{\geq}{\geqslant}
\renewcommand{\leq}{\leqslant}

\newcommand{\OPS}{\mathcal{O}_\text{PS}}
\newcommand{\OCDW}{\mathcal{O}_\text{CDW}}

\newcommand{\nn}{\nonumber}
\newcommand{\half}{{\textstyle{\frac{1}{2}}}}

\newcommand{\n}{\hat{n}}
\newcommand{\f}{\hat{f}}
\newcommand{\fd}{\hat{f}^\dag}

\DeclareMathOperator{\artanh}{artanh}
\allowdisplaybreaks

\begin{document}

\title{Off-diagonal long-range order and $p$-wave pairing in one dimension from critical density fluctuations}

\author{Matthew D. Thomson\,\orcidlink{0009-0004-5837-7430}}
\affiliation{Department of Physics, University of Alberta, Edmonton, Alberta T6G 2E1, Canada}
\affiliation{Theoretical Physics Institute \& Quantum Horizons Alberta, University of Alberta, Edmonton, Alberta T6G 2E1, Canada}

\author{Simon M. Martin\,\orcidlink{0009-0002-1064-3416}}
\affiliation{Department of Physics, University of Alberta, Edmonton, Alberta T6G 2E1, Canada}
\affiliation{Theoretical Physics Institute \& Quantum Horizons Alberta, University of Alberta, Edmonton, Alberta T6G 2E1, Canada}

\author{Frank Marsiglio\,\orcidlink{0000-0003-0842-8645}}
\affiliation{Department of Physics, University of Alberta, Edmonton, Alberta T6G 2E1, Canada}
\affiliation{Theoretical Physics Institute \& Quantum Horizons Alberta, University of Alberta, Edmonton, Alberta T6G 2E1, Canada}

\author{Joseph Maciejko\,\orcidlink{0000-0002-6946-1492}}
\affiliation{Department of Physics, University of Alberta, Edmonton, Alberta T6G 2E1, Canada}
\affiliation{Theoretical Physics Institute \& Quantum Horizons Alberta, University of Alberta, Edmonton, Alberta T6G 2E1, Canada}

\date{\today}

\begin{abstract}
A central theme in the theory of unconventional superconductivity is that Cooper pairing can be mediated by bosonic collective modes other than lattice vibrations. However, collective fluctuations in reduced dimensions can also have a destabilizing effect on superconductivity, by suppressing long-range order or by destroying Fermi-liquid quasiparticles in the normal state. In this work, we study a one-dimensional (1D) model of spinless fermions with local interactions that exhibits a continuous quantum phase transition from a Luttinger liquid to a phase-separated state. Combining field-theoretic arguments with numerical exact diagonalization and matrix-product-state methods, we find that the quantum critical point exhibits two-particle off-diagonal long-range order and $p$-wave pairing. Our work shows that critical density fluctuations at a phase-separation instability can act as pairing glue for unconventional superconductivity in 1D, overcoming the disordering effect of phase fluctuations and the absence of long-lived fermionic quasiparticles in the normal state.
\end{abstract}

\maketitle

\section{Introduction}

While conventional $s$-wave superconductivity is generally believed to arise from attractive effective interactions between electrons mediated by phonons~\cite{bardeen1957}, the nature of the pairing mechanism for unconventional (e.g., $p$-wave or $d$-wave) superconductors remains a largely open question. By analogy with the phonon mechanism of Bardeen-Cooper-Schrieffer (BCS) theory, a broad class of theories ascribe the formation of unconventional Cooper pairs to fermion interactions mediated by other bosonic collective modes, for example, ferromagnetic spin fluctuations in $p$-wave superfluid $^3$He~\cite{layzer1971,layzer1974,leggett1975} and antiferromagnetic spin fluctuations~\cite{cyrot1986,scalapino1986,miyake1986} in $d$-wave superconductors.

An interesting scenario in which soft bosonic collective modes naturally arise is the vicinity of a symmetry-breaking quantum critical point (QCP). At a QCP, long-wavelength fluctuations of the order parameter become gapless and can mediate interactions between electrons. Recent theoretical proposals along this line include pairing mediated by quantum-critical fluctuations of loop-current order~\cite{aji2010}, ferroelectricity~\cite{edge2015,wolfle2018,kozii2019}, and nematicity~\cite{lederer2015,lederer2017}. However, quantum-critical fluctuations can also result in a non-Fermi liquid and destroy the coherent fermionic quasiparticles in the normal state, an effect which competes with the Cooper instability~\cite{metlitski2015}. This effect is exacerbated in lower dimensions, and is particularly acute in one spatial dimension (1D), where interacting electrons generally form a Luttinger liquid~\cite{haldane1981} without coherent fermionic excitations. Furthermore, the hallmark of superconductivity, two-particle off-diagonal long-range order (ODLRO)~\cite{yang1962}, is naively forbidden in 1D even at zero temperature due to the Mermin-Wagner theorem~\cite{mermin1966,hohenberg1967,coleman1973}. Indeed, because of the disordering effect of long-wavelength fluctuations of the $U(1)$ phase of the superconducting order parameter, pairing correlations decay algebraically at long distances, suppressing true ODLRO in favor of quasi-ODLRO.

In this work, we study a 1D lattice model of spinless fermions with local interactions that exhibits $p$-wave pairing and true two-particle ODLRO at a QCP in its phase diagram. The QCP describes a continuous quantum phase transition from a Luttinger liquid to a phase-separated (PS) state with a spontaneously broken $\mathbb{Z}_2$ particle-hole symmetry. The order parameter for this transition corresponds to the uniform (zero-momentum) component of the fermion number/charge density relative to half filling, and critical fluctuations of this order parameter are key to the existence of superconducting ODLRO in this model. Conceptually, number and phase are canonically conjugate quantum-mechanical variables which do not mutually commute, thus strong density fluctuations at a QCP are naturally expected to lead to reduced fluctuations of the superconducting (Josephson) phase. More quantitatively, using bosonization techniques, we derive an effective field theory for the PS transition and show that it is equivalent to a previously studied field theory describing Ising ferromagnetic transitions in spin chains~\cite{nahum2025,flores-calderon2025,chepiga2026,shimomura2026,rey2026} and 1D spinful itinerant electron systems~\cite{yang2004,kozii2017,andriyakhina2026}. Importantly, Ref.~\cite{nahum2025} recently argued that this field theory exhibits spontaneous breaking of the continuous $U(1)$ symmetry associated with XY spin rotations around the Ising axis. In our work, the $U(1)$ symmetry corresponds instead to the conservation of fermion number, and its spontaneous breaking at the phase-separation QCP is associated with $p$-wave superconductivity. To verify the field theory predictions, we study the model numerically using exact diagonalization (ED) as well as matrix-product state (MPS) methods and confirm the existence of both the phase-separation QCP and true $p$-wave ODLRO at this QCP.

The rest of the paper is structured as follows. In Sec.~\ref{sec:model}, we present the model and discuss its symmetries and possible symmetry-breaking orders. In Sec.~\ref{sec:MFT-fluct}, we first show the existence of a PS critical point in this model using Hartree-Fock theory, and then derive a low-energy field theory for the transition using bosonization. In Sec.~\ref{sec:numerics}, we present detailed results for finite systems using ED as well as the density-matrix renormalization group (DMRG)~\cite{schollwock2011}, and directly in the thermodynamic limit using the variational uniform matrix product state algorithm (VUMPS)~\cite{zauner-stauber2018}. In addition to confirming the continuous nature of the PS transition and the presence of $p$-wave ODLRO at the QCP, we extract critical exponents for the transition and find good agreement with values previously reported~\cite{flores-calderon2025,chepiga2026,shimomura2026}. We also study the single-particle fermion Green's function at the QCP and find that it exhibits neither the exponential spatial decay of a BCS superconductor nor the power-law decay of a Luttinger liquid, but rather a \emph{stretched-exponential} decay as recently predicted in Ref.~\cite{andriyakhina2026}. We further argue that this implies that the single-particle momentum distribution $n(k)$ at the QCP is a \emph{smooth but nonanalytic} function at the Fermi wave vector $k=k_F$, a very weak type of singularity intermediate between the non-singular behavior predicted by BCS theory and the power-law singularity of Luttinger liquids. We conclude in Sec.~\ref{sec:conclusion} with a summary of our main results as well as an outline of possible directions for future research.

\section{Model Hamiltonian}
\label{sec:model}

\begin{figure}[t]
\includegraphics[width=\columnwidth]{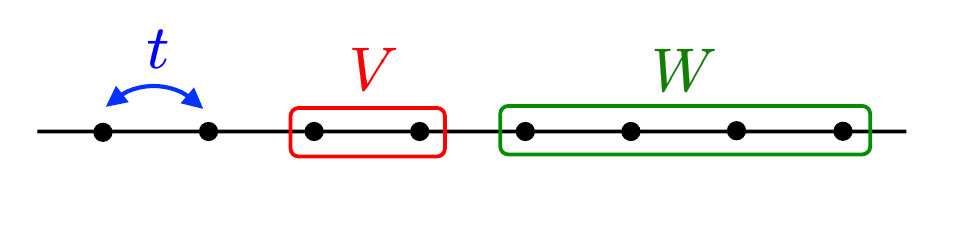}
\caption{Spinless fermions on a 1D lattice with hopping amplitude $t$, two-body density interaction $V$, and four-body density interaction $W$.}
\label{fig:model}
\end{figure}

We consider interacting spinless fermions described by the annihilation (creation) operators $\hat{f}_r$ ($\hat{f}_r^{\dag}$) on a 1D lattice with $L$ sites labeled by $r=0,\ldots,L-1$ and lattice constant $a$, which we set to unity unless otherwise specified. We define the Hamiltonian to be (see Fig.~\ref{fig:model}):
\begin{align}\label{Hfermion}
\hat{H}=\hat{H}_0+\hat{H}_2+\hat{H}_4,
\end{align}
where
\begin{align}
\hat{H}_0=-t\sum_r(\hat{f}_r^\dag\hat{f}_{r+1}+\mathrm{H.c.}),
\end{align}
is a nearest-neighbor hopping term;
\begin{align}\label{H2bdy}
\hat{H}_2=V\sum_r\delta\hat{n}_r\delta\hat{n}_{r+1},
\end{align}
is a nearest-neighbor density-density interaction; and
\begin{align}\label{H4bdy}
\hat{H}_4=W\sum_r\delta\hat{n}_r\delta\hat{n}_{r+1}\delta\hat{n}_{r+2}\delta\hat{n}_{r+3},
\end{align}
is a four-body interaction among fermions on four neighboring sites. Here, we define the lattice charge density relative to half filling:
\begin{align}\label{densityop}
\delta\hat{n}_r=\hat{n}_r-\half,
\end{align}
where $\hat{n}_r=\hat{f}_r^\dag\hat{f}_r$ is the number operator on site $r$. We will study the model for both periodic boundary conditions (PBC) and open boundary conditions (OBC).

\subsection{Symmetries}

The model (\ref{Hfermion}) has a continuous $U(1)$ symmetry generated by the total particle number
\begin{align}\label{Q}
\hat{Q}=\sum_r \hat{n}_r,
\end{align}
which commutes with the Hamiltonian. The model also has discrete symmetries. In addition to parity (inversion) $\hat{P}$ about the center of the chain,
\begin{align}\label{P}
\hat{P}\hat{f}_r\hat{P}^\dag=\hat{f}_{L-1-r},
\end{align}\label{T}
and translation symmetry $\hat{T}$ under PBC,
\begin{align}\label{T}
\hat{T}\hat{f}_r\hat{T}^\dag=\hat{f}_{r+1},
\end{align}
the Hamiltonian (\ref{Hfermion}) commutes with the particle-hole (PH) symmetry or charge-conjugation operator $\hat{C}$,
\begin{align}\label{C}
\hat{C}\hat{f}_r\hat{C}^\dag=(-1)^r \hat{f}_r^\dag,
\end{align}
which maps $\hat{C}\hat{n}_r\hat{C}^\dag=1-\hat{n}_r$. Both $\hat{P}$ and $\hat{C}$ are unitary $\mathbb{Z}_2$ symmetries; there is also an antiunitary $\mathbb{Z}_2$ time-reversal symmetry $\hat{\Theta}$ which reduces to complex conjugation (spinless time reversal) and will play no role in our discussion.

We will be interested in the ground-state physics of the Hamiltonian (\ref{Hfermion}), and the PH symmetry $\hat{C}$ will play an important role. Since it is a discrete symmetry, it can be spontaneously broken in 1D at zero temperature. If the PH symmetry is unbroken in the ground state $\ket{0}$, i.e., $\hat{C}\ket{0}=\eta\ket{0}$ where $\eta=\pm 1$, then $\langle \hat{n}_r\rangle=\half$ for all $r$, where $\langle\cdots\rangle$ denotes the ground-state expectation value. By contrast, if the PH symmetry is spontaneously broken, then there are (at least) two degenerate ground states related by $\hat{C}$, and in either of those ground states, $\langle\hat{n}_r\rangle$ will deviate from $\half$. The ground-state expectation value $\langle\delta\hat{n}_r\rangle$ of the operator (\ref{densityop}) can thus be taken as a local order parameter for the spontaneous breakdown of PH symmetry. For simplicity we refer below to $\delta\hat{n}_r$ as the density operator.

\subsection{Charge-density wave (CDW) order vs phase separation (PS)}

If PH symmetry is broken spontaneously, two further scenarios are possible. First, the breakdown of PH symmetry can be accompanied by a spontaneous breakdown of translation symmetry. For repulsive density-density interactions, this is the likely scenario, i.e., a charge-density wave (CDW) ground state where $\langle \delta\hat{n}_r\rangle\propto(-1)^r$ and the unit cell is doubled. Alternatively, the breakdown of PH symmetry can occur in the \emph{absence} of translation symmetry breaking (absent domains), which corresponds to a phase-separation (PS) or segregation instability~\cite{freericks1990,emery1990}. This scenario is favored by attractive density-density interactions. While the global (averaged) order parameter for CDW order is
\begin{align}
\OCDW=\frac{1}{L}\sum_r(-1)^r\langle \delta\hat{n}_r\rangle,
\end{align}
that for PS is
\begin{align}\label{OPS}
\OPS=\frac{1}{L}\sum_r\langle\delta\hat{n}_r\rangle=n-\half,
\end{align}
where
\begin{align}\label{n}
n=\frac{\langle\hat{Q}\rangle}{L}
\end{align}
is the \emph{equilibrium} density, i.e., the density corresponding to the state of lowest energy. A CDW instability corresponds to two degenerate ground states with $\mathcal{O}_\text{CDW}=\pm m\neq 0$ but $\mathcal{O}_\text{PS}=0$, while a PS instability corresponds to two degenerate ground states with $\mathcal{O}_\text{PS}=\pm m\neq 0$ but $\mathcal{O}_\text{CDW}=0$. Put differently, in a CDW instability the lowest energy is found at half filling, $n=\half$, while in a PS instability the lowest energy is found at two densities $n$ and $1-n$ related by PH symmetry but \emph{not} equal to $\half$. If the PS instability occurs while the total particle number is held fixed, e.g., at half filling, the system will spontaneously form Ising-like domains corresponding to the two different equilibrium values $n$ and $1-n$.

Below, we will show that the model (\ref{Hfermion}) exhibits a PS instability at zero temperature in a certain range of couplings, and we will find that exactly at the quantum phase transition between a uniform ground state and a phase-separated ground state, the model exhibits true ODLRO, i.e., ($p$-wave) superconductivity, despite being in 1D.

\section{Mean-field theory \& Fluctuations}
\label{sec:MFT-fluct}

We begin by studying the Hamiltonian (\ref{Hfermion}) using mean-field theory. In Sec.~\ref{Sec:hartree}, we use a simple Hartree approximation to show that the model exhibits a continuous PS instability at a finite critical value $g_c$ of the coupling $g\equiv V/t$ for fixed $W/t>0$ (at least when exchange terms are excluded). In Sec.~\ref{Sec:HF}, we refine this analysis to include all exchange terms, i.e., the full Hartree-Fock approximation, and show that the transition persists and remains continuous. In Sec.~\ref{Sec:EFT}, we go beyond mean-field theory and derive a low-energy effective field theory for the PS transition via Abelian bosonization.

\subsection{Hartree approximation}\label{Sec:hartree}

\begin{figure}[t]
\includegraphics[width=\columnwidth]{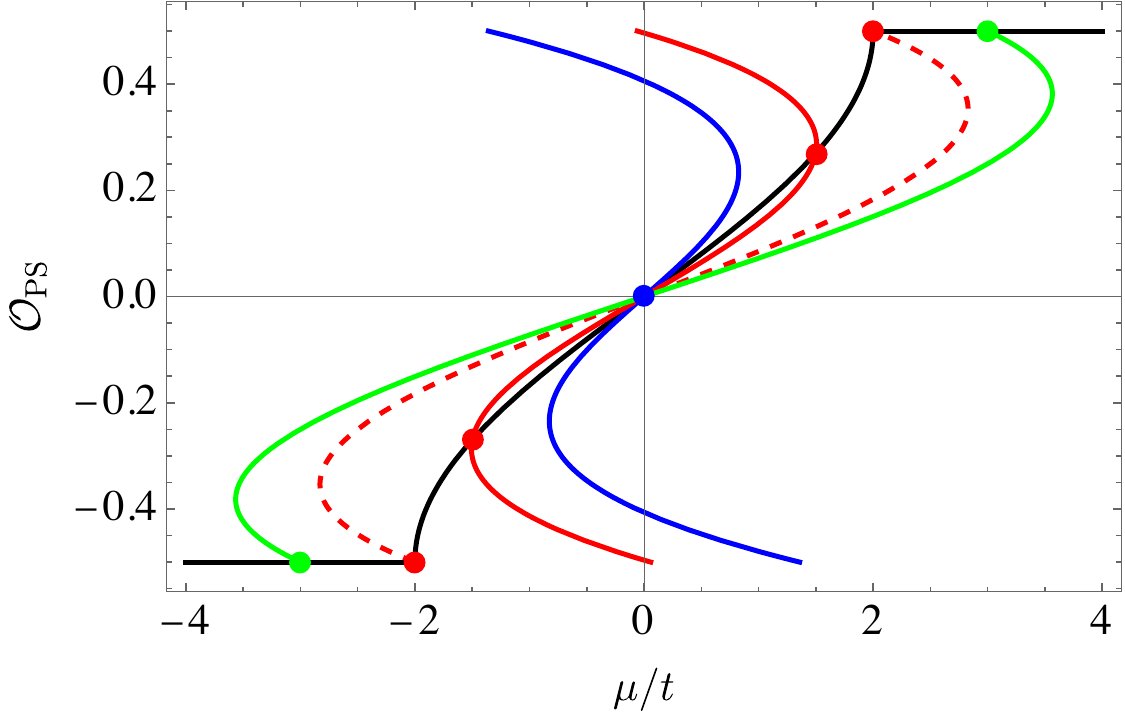}
\caption{PS instability in the Hartree approximation. The black curve represents the PS order parameter $\OPS$ obtained from the number equation (\ref{gapequation}), while the colored curves are plots of the self-consistency condition (\ref{mudef}) for sufficiently large $W>0$ and different values of $g\equiv V/t$. For $g>g_c$, only the symmetric solution $\OPS=0$ is found (blue dot), while for $g<g_c$, a pair of broken-symmetry solutions appears at $\OPS=\pm m\neq 0$ (red and green dots). The red dashed line corresponds to $g=g_*<g_c$, below which the order parameter saturates to its maximal value $\pm\half$ (green dots).}
\label{fig:hartree}
\end{figure}

In the Hartree approximation, which ignores exchange terms coming from particle indistinguishability, we write the density operator (\ref{densityop}) as $\delta\hat{n}_r=\OPS+\delta\c{\hat{O}}_\text{PS}(r)$ and expand the interaction terms (\ref{H2bdy}-\ref{H4bdy}) to linear order in the density fluctuations $\delta\c{\hat{O}}_\text{PS}(r)=\delta\hat{n}_r-\OPS$. The resulting Hartree (H) Hamiltonian is
\begin{align}\label{HFHamiltonian}
\hat{H}_\text{H}=\sum_k(\varepsilon_k-\mu)\hat{f}_k^\dag\hat{f}_k+\text{const.},
\end{align}
where $\varepsilon_k=-2t\cos k$ is the single-particle dispersion and
\begin{align}\label{mudef}
\mu=-2V\OPS-4W\OPS^3,
\end{align}
is the chemical potential. The order parameter $\OPS$, or equivalently the chemical potential $\mu$, must obey a self-consistency condition, which is none other than the number equation
\begin{align}
\OPS=-\frac{1}{2}+\int_{-\pi}^\pi\frac{dk}{2\pi}\langle\hat{f}^\dag_k\hat{f}_k\rangle_\text{H},
\end{align}
where the expectation value is taken in the ground-state of (\ref{HFHamiltonian}), i.e., the free Fermi sea with chemical potential $\mu$. One thus obtains
\begin{align}\label{gapequation}
\OPS=\left\{
\begin{array}{rr}
-\frac{1}{2}+\frac{1}{\pi}\cos^{-1}\left(-\frac{\mu}{2t}\right), & |\mu|<2t, \\
\half\sgn\mu, & |\mu|>2t,
\end{array}\right.
\end{align}
which is to be solved self-consistently with Eq.~(\ref{mudef}).

The self-consistent Hartree equations (\ref{mudef}) and (\ref{gapequation}) can be solved graphically by plotting them together and looking for points of intersection (Fig.~\ref{fig:hartree}). As $g$ is decreased for fixed $W/t>0$, we find a continuous quantum phase transition at $g=g_c$ such that for $g>g_c$, only the trivial solution $\OPS=0$ exists, whereas for $g<g_c$, there also exists a pair of broken-symmetry solutions $\OPS=\pm m\neq 0$.

To verify that the transition is continuous, and to find $g_c$, we assume that the order parameter $\OPS$ is small near the critical point and expand the number equation,
\begin{align}\label{NumEqExpand}
\mu=-2t\cos[\pi(\OPS+\half)]=2\pi t\OPS-\frac{\pi^3t}{3}\OPS^3+\ldots
\end{align}
Equating Eqs.~ (\ref{mudef}) and (\ref{NumEqExpand}), we find the nontrivial solution as
\begin{align}
\OPS=\pm\sqrt{-\frac{g+\pi}{(2W/t)-\pi^3/6}},
\end{align}
for $g<-\pi$. Thus, the Hartree approximation predicts a continuous phase-separation instability at $g_c=-\pi$, provided that the strength of four-body interactions is sufficiently large, $W/t>\pi^3/12\approx 2.58$. For sufficiently negative $g$, the order parameter saturates to its maximal value, which happens below $g_*=-2-W/(2t)$.

\subsection{Hartree-Fock approximation}\label{Sec:HF}

\begin{figure}[t]
\includegraphics[width=\columnwidth]{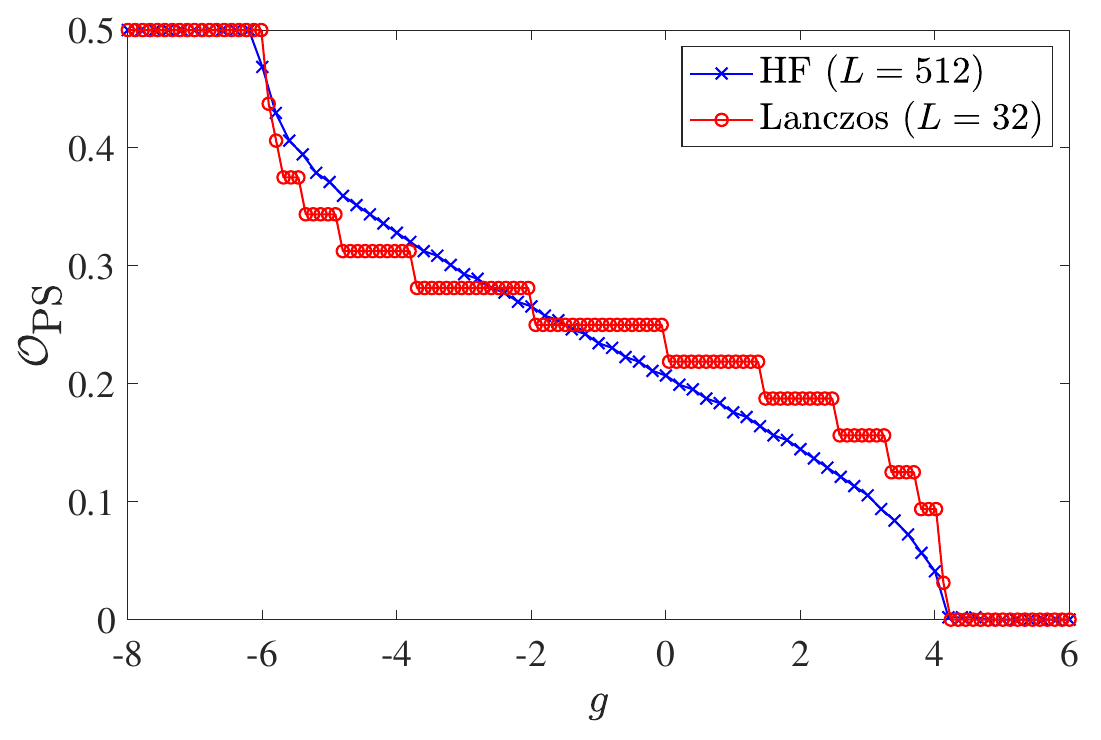}
\caption{PS order parameter computed as a function of $g\equiv V/t$ in the HF approximation (blue curve) for $L=512$ sites, and using Lanczos ED (red curve) for $L=32$ sites. Here $W/t = 8$ and PBC are used in both cases.}
\label{fig:HF-ED}
\end{figure}

The Hartree approximation provides an analytically tractable, simple conceptual picture of the PS instability but ignores exchange energy contributions which can be substantial in Fermi systems. To verify that the transition persists when exchange terms are accounted for, we compute the ground-state energy density $\varepsilon_\text{GS}(n)$ in the full Hartree-Fock (HF) approximation,
\begin{align}\label{eGS-HF}
\varepsilon_\text{GS}(n)=\frac{\langle\hat{H}\rangle}{L}.
\end{align}
The expectation value of the Hamiltonian is taken in a free Fermi sea with density $n$ and thus chemical potential $\mu=-2t\cos(\pi n)$, and we take all nonzero Wick contractions into account. Thus, as is well known~\cite{mahan}, the HF approximation is equivalent to first-order perturbation theory in the interaction strength (here, the couplings $V$ and $W$). Analytical expressions for the different terms entering the right-hand side of Eq.~(\ref{eGS-HF}) are given in Appendix~\ref{app:HF} as momentum sums over the first Brillouin zone, together with their diagrammatic representations. In practice, to approximate the thermodynamic limit $L\rightarrow\infty$, we perform the sums numerically for finite $L$ and find that the HF energy density $\varepsilon_\text{GS}(n)$ does not vary appreciably beyond $L$ on the order of a few hundred sites. Particle-hole symmetry $\varepsilon_\text{GS}(1-n)=\varepsilon_\text{GS}(n)$ is verified up to numerical accuracy, thus we restrict ourselves to $\half\leq n\leq 1$. Within this range, we find numerically the equilibrium density that minimizes $\varepsilon_\text{GS}(n)$, and from it we compute the PS order parameter (\ref{OPS}).

The PS order parameter computed from the HF approximation is plotted in Fig.~\ref{fig:HF-ED} (blue curve) for $W/t=8$. Consistent with the results of the Hartree approximation, we find a continuous PS transition, but exchange corrections significantly renormalize the position of the critical point ($g_c\approx 4.2$). Another continuous transition to the fully phase-separated state ($\OPS=\pm 0.5$) occurs around $g_*\approx-6.2$.

We note here that the four-body term (\ref{H4bdy}) is important to obtain a continuous PS transition. Indeed, for $W=0$, our Hamiltonian (\ref{Hfermion}) reduces to the $t$-$V$ model, which is exactly solvable by the Bethe ansatz and exhibits a single first-order transition into the fully phase-separated state at $g_c=g_*=-2$~\cite{schollwock2004,nakamura1997,biswas2026}.

\subsection{Bosonization \& low-energy field theory}\label{Sec:EFT}

Since Hartree-Fock mean-field theory ignores the effects of quantum fluctuations, which are crucial in 1D, we next use Abelian bosonization~\cite{giamarchi2003} to derive a low-energy continuum field theory describing the system near the PS transition, as it is approached from the disordered phase. The theory is first written in terms of the two periodic scalar fields $\phi$ and $\theta$, where $\phi$ is related to the smooth part of the charge density relative to half filling $\delta\rho(x) = \rho(x)-\rho_0$, with $\rho_0 = \frac{1}{2a}$, and we have restored the lattice spacing $a$. In the bosonized description, the mapping is
\begin{equation}\label{PS-OP-bos}
\frac{1}{a} \delta \hat{n}_r \rightarrow \delta \rho(x) \approx -\frac{1}{\pi} \partial_x\phi ,
\end{equation}
where $x = a r$. On the other hand, the Josephson phase $\theta$ is related to $\phi$ via the canonical commutation relation $[\phi(x),\frac{1}{\pi} \partial_{x'} \theta(x')] = i \delta(x-x')$. We detail in Appendix~\ref{Sec:Bosonization_H} the Abelian bosonization of the fermionic Hamiltonian $\hat{H}$ in Eq.~(\ref{Hfermion}), which results in the following low-energy continuum Hamiltonian:
\begin{align}\label{Eq:Bosonized_H_main}
\begin{split}
\hat{H} &= \frac{1}{2\pi} \int dx \, \Big[ u K (\partial_x \theta)^2 + \frac{u}{K} (\partial_x \phi)^2 \Big] \\ &\hspace{0.5cm} + \int dx \Big[ g_1 (\partial_x^2 \phi)^2 + g_2 (\partial_x \phi)^4 + g_3 \cos(4\phi) \Big] + \dots .
\end{split}
\end{align}
Here, the first line corresponds to the usual Tomonaga-Luttinger Hamiltonian~\cite{giamarchi2003}, with $u$ and $K$ corresponding respectively to the renormalized Fermi velocity and the Luttinger parameter. The second line corresponds to interactions and/or terms arising at higher orders in the gradient expansion, with their corresponding couplings $g_1$, $g_2$, and $g_3$, while the ellipsis denotes terms that are even more irrelevant\footnote{Some of these higher-order terms however contribute to the renormalization of leading terms of the low-energy Euclidean action, as detailed in Appendix~\ref{Sec:Bosonization_H}.}. The cosine interaction corresponds to the umklapp process allowed for spinless fermions at commensurate filling (half filling in our case). The parameters $u$, $K$, $g_1$, $g_2$, and $g_3$ in the field theory have bare values determined by the microscopic parameters $a$, $t$, $V$, and $W$ (Appendix~\ref{Sec:Bosonization_H}), but they flow under the renormalization group and are thus kept simply as low-energy couplings parameterizing the continuum Hamiltonian.

The corresponding Euclidean action is obtained by performing a Legendre transformation. Integrating out the Gaussian fluctuations of $\theta$ results in the following low-energy theory, after an appropriate rescaling (Appendix~\ref{Sec:Bosonization_H}):
\begin{align}\label{Eq:Bosonized_S}
\begin{split}
S[\phi] &= \int d\tau \, dx \Bigg[ \frac{1}{2} (\partial_{\tau} \phi)^2 + \frac{1}{2} (\partial_x^2 \phi)^2 \\ &\hspace{2cm} + \lambda_2 (\partial_x \phi)^2 + \lambda_4 (\partial_x \phi)^4 \Bigg] + \dots .
\end{split}
\end{align}
The local order parameter for the PS transition is precisely the density relative to half filling (\ref{PS-OP-bos}). Therefore, the transition occurs when the coefficient $\lambda_2$ vanishes (at least at tree-level), or equivalently when $K \rightarrow \infty$ in Eq.~(\ref{Eq:Bosonized_H_main}). We provide numerical evidence of this feature in Sec.~\ref{sec:numerics}. Since the scaling dimension of the umklapp term is $\Delta_{\cos(4\phi)} = 4K$ in the Luttinger liquid phase~\cite{giamarchi2003}, it becomes strongly irrelevant as the critical point is approached and can thus be dropped. Note also the importance of keeping the two higher-order derivative terms $(\partial_x^2\phi)^2$ and $(\partial_x\phi)^4$, where the former is the leading spatial gradient term at criticality, while the latter ensures the thermodynamic stability of the theory in the broken-symmetry phase ($\lambda_2<0$).

The Lifshitz-like field theory (\ref{Eq:Bosonized_S}) is identical in form to a previously studied field theory describing continuous Ising ferromagnetic transitions in spin chains~\cite{nahum2025,flores-calderon2025,chepiga2026,shimomura2026,rey2026} and 1D spinful itinerant electron systems~\cite{yang2004, kozii2017,andriyakhina2026}. (The Gaussian limit of this theory, i.e., the first two terms in Eq.~(\ref{Eq:Bosonized_S}), was also considered in Refs.~\cite{radzihovsky2023,kuklov2024}.) Recently, some of these works \cite{nahum2025,flores-calderon2025,chepiga2026,andriyakhina2026,shimomura2026,rey2026} have shown that exactly at the critical point between the Ising paramagnetic ($\frac{1}{L}\sum_r\langle\hat{S}^z_r\rangle=0$) and the Ising ferromagnetic ($\frac{1}{L}\sum_r\langle\hat{S}^z_r\rangle\neq 0$) phases, spontaneous breaking of a continuous $U(1)$ symmetry under $\hat{S}^z$ spin rotation occurs. In spin-chain models, this manifests itself as XY ferromagnetic long-range order ($\frac{1}{L}\sum_r\langle\hat{S}^+_r\rangle\neq 0$), whereas in 1D spinful itinerant electron systems, it instead manifests itself as spin-nematic long-range order $(\frac{1}{L}\sum_r\langle\hat S_r^+\hat S_{r+1}^+\rangle\neq0)$ \cite{andriyakhina2026}. By contrast, in our model, the $U(1)$ symmetry is generated by the total particle number (\ref{Q}), but the operator most analogous to $\hat{S}_r^+$ (i.e., with the same $U(1)$ charge) is the fermion creation operator $\hat{f}_r^\dag$ which cannot acquire a nonzero expectation value. However, $U(1)$ symmetry breaking in the present problem suggests the possibility of \emph{two-particle} ODLRO, i.e., Cooper pairing and superconductivity.

For spinless fermions, the natural superconducting order parameter on the lattice is the uniform $p$-wave pairing order parameter,
\begin{align}
    \Delta=\frac{1}{L}\sum_r\langle\hat{\Delta}_r\rangle,
\end{align}
where
\begin{align}\label{pairingOP}
    \hat{\Delta}_r=\hat{f}_r\hat{f}_{r+1},
\end{align}
which after bosonization (Appendix~\ref{Sec:Bosonization_SC_order_parameter}) results in
\begin{align}
    \hat{\Delta}_r \rightarrow \Delta(x) \sim e^{i 2 \theta(x)}.
\end{align}
In the Luttinger liquid phase ($\lambda_2>0$), the two-point correlation function of the superconducting order parameter decays as
\begin{align}\label{PairingCorrelatorBos}
 \langle \Delta^\dag(0) \Delta(x) \rangle \sim 1/|x|^{2/K}.   
\end{align}
Hence, as the PS critical point is approached, $K$ diverges and the two-point function decays more and more slowly. Exactly at the critical point ($K=\infty$), we expect true ODLRO:
\begin{align}
    \lim_{x\rightarrow\infty}\langle \Delta^\dag(0) \Delta(x) \rangle =|\Delta|^2=\text{const.}
\end{align}

To verify those expectations from the bosonized field theory, we now turn to numerical calculations using exact diagonalization (ED), the density-matrix renormalization group (DMRG), and the variational uniform matrix product state algorithm (VUMPS).

\section{Numerical results}\label{sec:numerics}

\subsection{Exact diagonalization (ED)}

We first study the Hamiltonian (\ref{Hfermion}) using exact diagonalization (ED). As the Hamiltonian preserves total particle number $\hat{Q}$ [Eq.~(\ref{Q})], we work in sectors of fixed quantum number $Q$ and find the absolute ground state by selecting the state of lowest energy across all sectors. Because of the particle-hole symmetry (\ref{C}), the $Q$ and $L-Q$ sectors are exactly degenerate, thus it is sufficient to restrict ourselves to $\frac{L}{2}\leq Q\leq L$. We use periodic boundary conditions (PBC) and implement translation symmetry $\hat{T}$ [Eq.~(\ref{T})] by working in the momentum ($k$) basis. We first study the model using full ED for system sizes $L=12,16,20$, setting $W/t=8$ and varying $g$ as in Fig.~\ref{fig:HF-ED}. We find that the absolute ground state is in the $k=0$ sector in all cases. We use Lanczos diagonalization for larger system sizes ($L=24,28,32$), fixing the momentum sector to be $k=0$ and implementing spatial parity $\hat{P}$ [Eq.~(\ref{P})] to further reduce the Hilbert space size. The absolute ground state is found to be in the $P=+1$ sector in all cases.

\begin{figure}[t]
\includegraphics[width=\columnwidth]{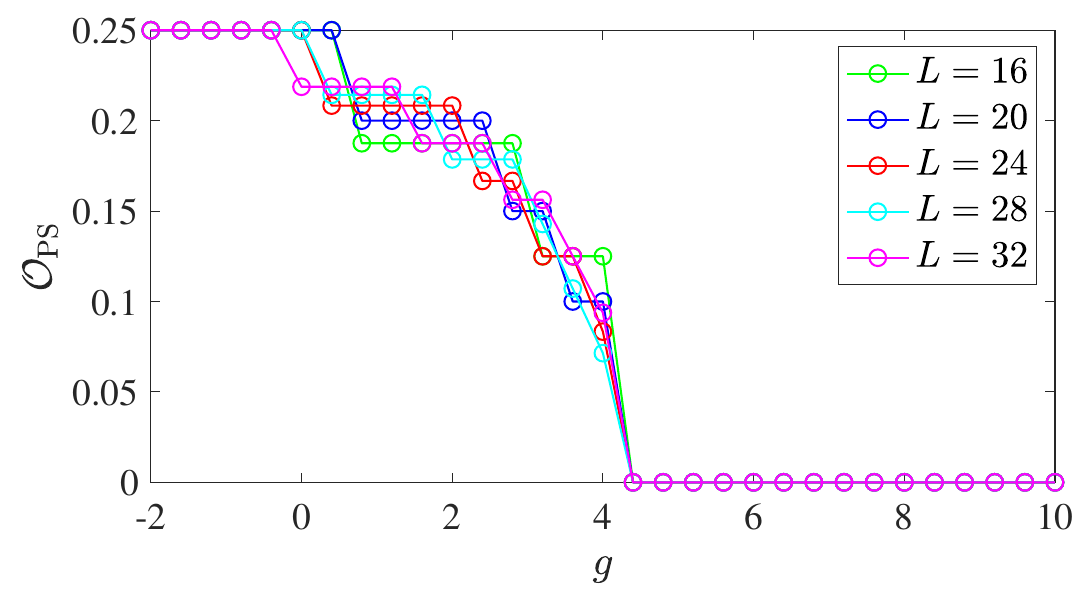}
\caption{PS order parameter computed using Lanczos ED for $W/t=8$, for a chain of length $L$ with PBC.}
\label{fig:PS-ED}
\end{figure}

The PS order parameter $\OPS=n-\half$ is obtained directly from the ground-state density $n=Q/L$. It is plotted in Fig.~\ref{fig:PS-ED} for various system sizes in the vicinity of the PS transition, and also in Fig.~\ref{fig:HF-ED} (red curve) for $L=32$ over a wider range of values of $g$. The phase boundaries obtained from the mean-field HF calculation are in remarkable agreement with the ED result, and the order parameter curve depends only weakly on the system size. The step-like nature of the PS order parameter owes to the fact that the conserved charge $Q$ is an integer and thus the density $n=Q/L$ necessarily assumes discrete values in a finite system. Nevertheless the PS transition appears continuous, although the precise shape of the equilibrium density curve differs significantly from the mean-field results, as expected in 1D. 

Having established the existence of a continuous PS transition in the fermionic Hamiltonian (\ref{Hfermion}) via both mean-field and ED approaches, we next turn to matrix-product state (MPS) methods~\cite{schollwock2011} for a more quantitative description of its critical properties, including the emergence of $p$-wave ODLRO as predicted by the field-theory arguments of Sec.~\ref{Sec:EFT}.

\subsection{Density-matrix renormalization group (DMRG)}

\begin{figure}[t]
    \includegraphics[width=\columnwidth]{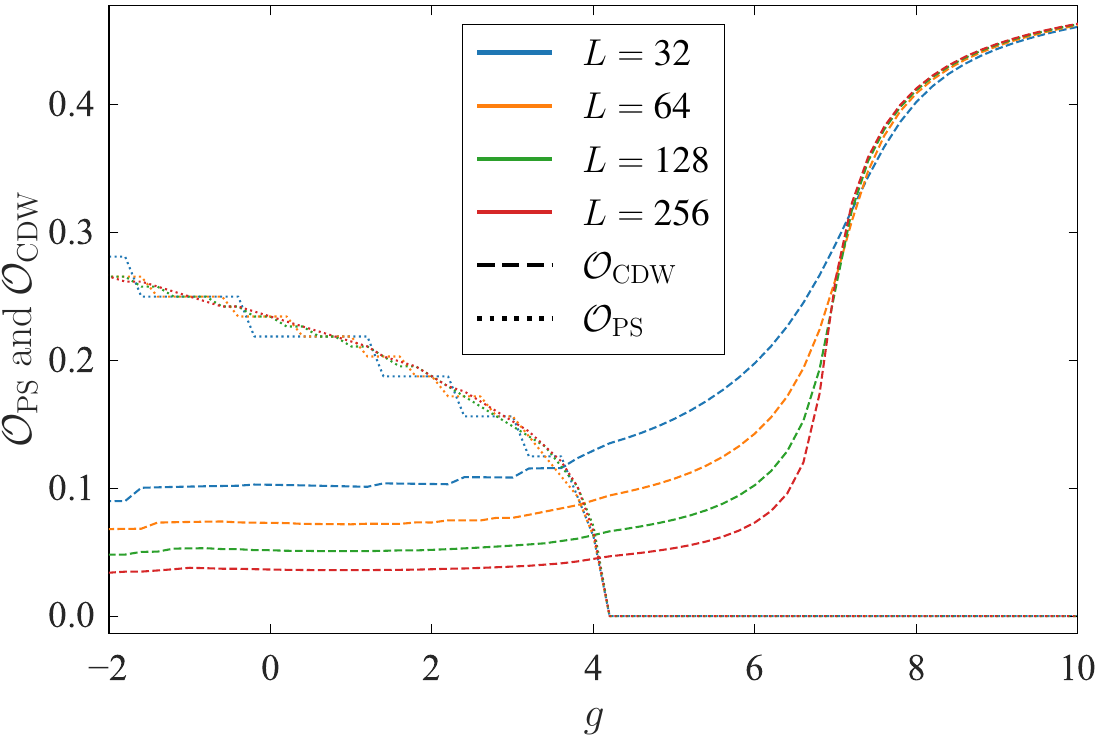}
    \caption{PS and CDW order parameters from finite-size OBC DMRG calculations at $W/t=8$. The bond dimension was fixed at $\chi=600$.}
    \label{fig:O_PS_O_CDW}
\end{figure}

We first compute the ground state at finite system sizes under OBC using the ITensor implementation of DMRG~\cite{fishman2022}. The DMRG methods are explained in more detail in Appendix \ref{sec:DMRG}. Figure~\ref{fig:O_PS_O_CDW} shows both the PS ($q=0$) and CDW ($q=\pi$) order parameters over the same range of $g$ as in Fig.~\ref{fig:PS-ED}, computed from the square root of the ground-state equal-time structure factor,
\begin{align}
    S_{\delta n}(q)=\frac{1}{L^2}\sum_{r,r'=0}^{L-1}e^{-iq(r-r')}\langle\delta\hat{n}_r\delta\hat{n}_{r'}\rangle.
    \label{deltan_structure_factor}
\end{align}
As similarly seen in the ED results, the quantization of the PS order parameter decreases with the system size as the number of fixed $Q$ sectors increases. The larger system sizes accessible in DMRG allow us to confirm the continuous nature of the PS transition.

The CDW order parameter $\OCDW$ is non-zero for all $g$ but decreases steadily with system size for $g\lesssim 7$. By contrast, for $g\gtrsim 7$ we find that $\OCDW$ is largely independent of system size. Thus, there is good evidence for an intermediate Luttinger-liquid phase without any broken symmetry between the PS transition at $g_c\sim 4$ and a CDW transition at $g_c\sim 7$. We conjecture that this phase is adiabatically connected to the Luttinger liquid in the $t$-$V$ model for $|g|\leq 2$, to which Eq.~(\ref{Hfermion}) reduces with $W=0$, and thus that the CDW transition, which occurs in the $t$-$V$ model for $g=2$, is likewise also in the Kosterlitz-Thouless universality class~\cite{schollwock2004}.

\subsection{Variational uniform matrix product states (VUMPS)}

To study the model (\ref{Hfermion}) without the influence of finite size or edge effects, we use the VUMPS algorithm~\cite{zauner-stauber2018} to compute the ground state directly in the thermodynamic limit. The variational choice of ground state used is a translationally invariant uniform MPS. Details of the VUMPS implementation are presented in Appendix~\ref{sec:VUMPS}.

\begin{figure}[t]
\includegraphics[width=\columnwidth]{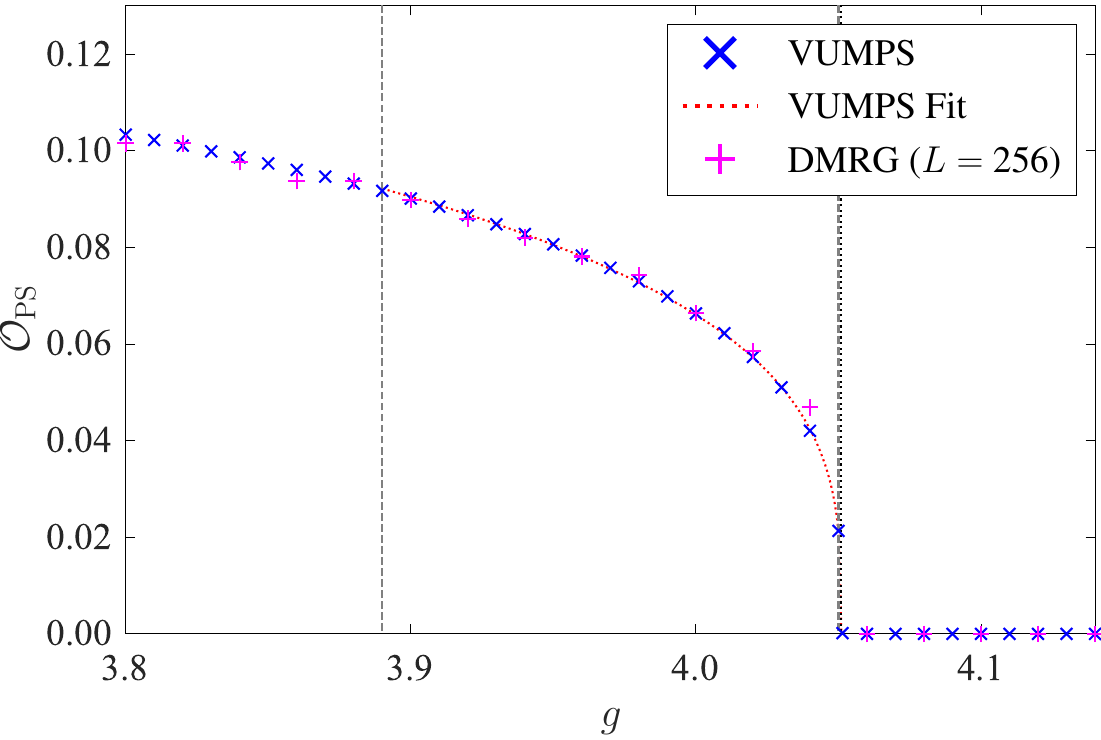}
\caption{PS order parameter computed from DMRG and VUMPS for $W/t=8$. VUMPS results were calculated with a bond dimension $\chi=1600$, using Eq.~(\ref{OPS2-VUMPS}). DMRG results were calculated with a bond dimension $\chi=600$ and system size $L=256$ under OBC, using the square root of Eq.~(\ref{deltan_structure_factor}). The red curve was fit to the VUMPS data only, with grey dashed lines indicating the range of $g$ considered in the fit. The dotted black line indicates the fit-optimized critical value of $g_{c,\beta}\approx 4.0511$.}

\label{fig:beta}
\end{figure}

First, we compute the PS order parameter from the expectation value of the density operator according to Eq.~(\ref{OPS}). Due to translation invariance of the uniform MPS and tendency for the VUMPS algorithm to converge to a maximally symmetry-broken state \cite{zauner-stauber2018}, avoiding highly-entangled linear combinations of the degenerate PS ground states (cat states), we can write the PS order parameter as:
\begin{align}\label{OPS2-VUMPS}
    \OPS=\langle \delta \hat{n}_0 \rangle,
\end{align}
where $\langle \delta \hat{n}_0 \rangle$ is the density expectation value at the central site of the uniform MPS. In Fig.~\ref{fig:beta}, we plot the absolute value of Eq.~(\ref{OPS2-VUMPS}) in the vicinity of the PS transition, computed with VUMPS with a bond dimension of $\chi=1600$, compared against DMRG results for $L=256$ and bond dimension $\chi=600$. By fitting the VUMPS order parameter curve to the power-law dependence
\begin{align}
    \OPS\sim(g_c-g)^\beta,\hspace{10mm}g<g_c,
\end{align}
 and optimizing the fit deviance against $g_c$ giving an estimated $g_{c,\beta}\approx4.0511$, we find the order-parameter critical exponent 
\begin{align}\label{beta}
    \beta=0.29\pm 0.02,
\end{align}
where we have estimated the uncertainty based on the sensitivity of the fit to the critical value. This value of $\beta$, distinct from both the mean-field value ($\beta=1/2$) and the standard (1+1)D Ising universality class ($\beta=1/8$), is in good agreement with the theoretical value $\beta=0.301$ obtained from a 2-loop field-theoretic renormalization group (RG) analysis~\cite{flores-calderon2025}, as well as values obtained from other DMRG studies~\cite{flores-calderon2025,chepiga2026,shimomura2026}.

\begin{figure}[t]
\includegraphics[width=\columnwidth]{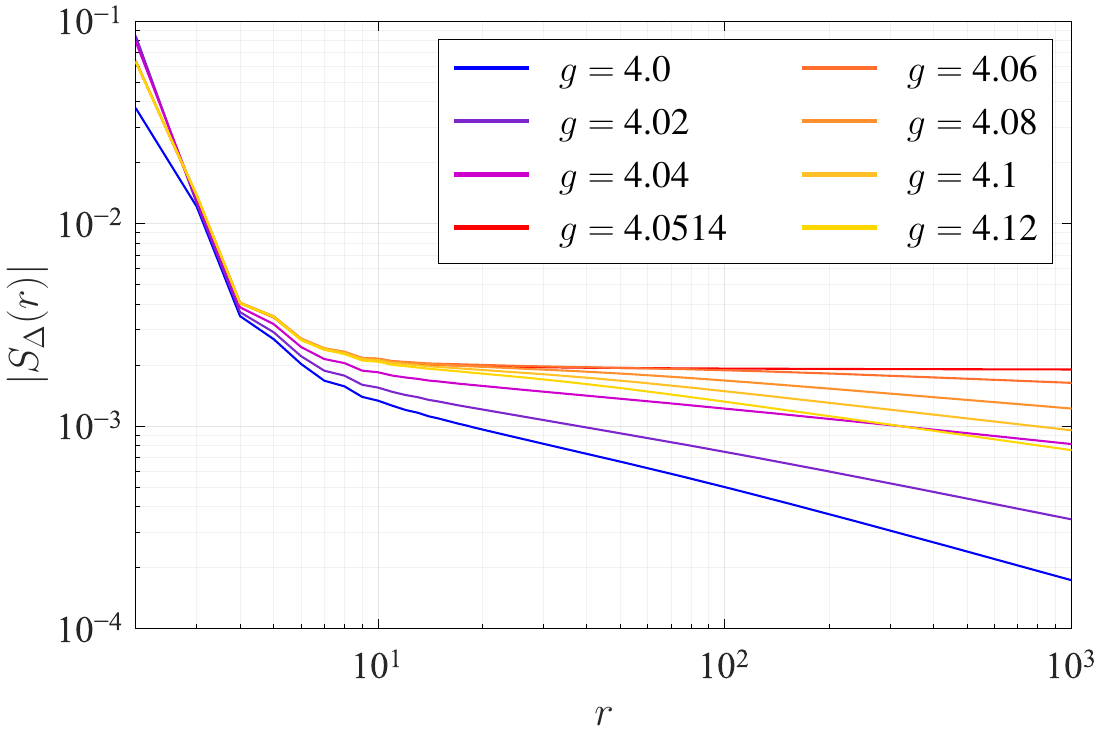}
\caption{Absolute value $|S_\Delta(r)|$ of the equal-time $p$-wave pairing correlator $S_\Delta(r)=\langle\hat{\Delta}^\dag_0\hat{\Delta}_r\rangle$ as a function of distance $r$, computed from VUMPS for $W/t=8$ with a bond dimension $\chi=1600$. At the critical point, the exponent of the power-law decay decreases to zero and the pairing correlator asymptotes to a nonzero constant at long distances, demonstrating two-particle ODLRO. Not all computed values of $g$ are shown for clarity.}
\label{fig:PairingCorrelator}
\end{figure}

Next, we investigate the equal-time two-point correlations $S_\Delta(r)\equiv\langle\hat{\Delta}^\dag_0\hat{\Delta}_r\rangle$ of the $p$-wave pairing order parameter (\ref{pairingOP}). In Fig.~\ref{fig:PairingCorrelator}, we plot the absolute value $|S_\Delta(r)|$ for several values of $g$ across the PS critical point $g_c$. We find that the pairing correlator asymptotes to a constant nonzero value at $g=g_c$, demonstrating two-particle $p$-wave ODLRO enabled by quantum critical density fluctuations, which is one of the main results of our work.

\begin{figure}[t]
\includegraphics[width=\columnwidth]{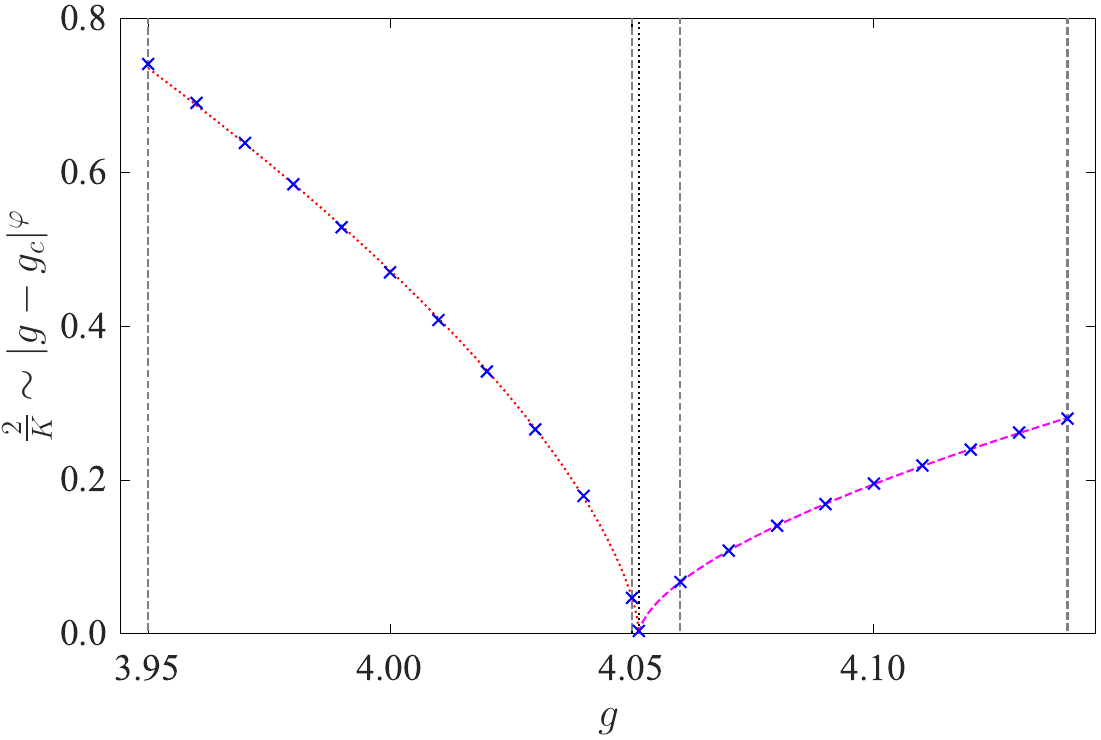}
\caption{Exponent $2/K$ characterizing the decay of $p$-wave pairing correlations [Eq.~(\ref{PairingCorrelatorBos})], obtained from power-law fits of $|S_\Delta(r)|$ as shown in Fig.~\ref{fig:PairingCorrelator} over a range of $r\in[300,800]$ to avoid short-distance effects and the finite correlation length imposed by the finite MPS bond dimension. The range of data used in the power-law fits for $\varphi=\nu(z-1)$ are marked with vertical dashed lines, and our best-guess estimate for $g_c=4.0514$ is excluded from both. The black dotted line marks the fit-optimized critical point $g_{c,\varphi}\approx 4.0515$ that both fits agree on. }
\label{fig:PairingCorrelatorLL}
\end{figure}

To make contact with the bosonization analysis in Sec.~\ref{Sec:EFT}, we fit the asymptotic decay of $|S_\Delta(r)|$ in Fig.~\ref{fig:PairingCorrelator} to the power-law form of Eq.~(\ref{PairingCorrelatorBos}). In Fig.~\ref{fig:PairingCorrelatorLL}, we plot the exponent $2/K$ obtained from this fit. On either side of the PS transition, the exponent $2/K$ is nonzero, consistent with quasi-ODLRO expected in a Luttinger liquid with $K<\infty$. At the PS transition however, we find that $2/K\rightarrow 0$, consistent with the identification of the PS quantum critical point with the divergence $K\rightarrow\infty$ of the Luttinger parameter in the bosonized field theory of Sec.~\ref{Sec:EFT}. We find that the Luttinger parameter diverges upon approach to the critical point as a power law, which was also observed in Ref.~\cite{chepiga2026} at the Ising ferromagnetic transition in a spin-1 chain and argued therein to scale as
\begin{align}
    \frac{1}{K}\sim|g-g_c|^\varphi,
\end{align}
where the crossover exponent is $\varphi=\nu(z-1)$. By fitting the data in Fig.~\ref{fig:PairingCorrelatorLL} and estimating the uncertainty based on the sensitivity of the fits to the critical value $g_{c,\varphi}$, we obtain an estimated exponent shared by both fits:
\begin{align}\label{phi}
\varphi= 0.63\pm 0.04,
\end{align}
which is in reasonable agreement with the 2-loop field-theoretical prediction $\varphi=0.592$~\cite{flores-calderon2025} and other DMRG results~\cite{chepiga2026}.

The power-law fits in Fig.~\ref{fig:PairingCorrelatorLL} 
were done similarly to Fig.~\ref{fig:beta}. The data with $g<g_c$ and $g>g_c$ both fit to the same critical value $g_{c,\varphi}\approx4.0515$. Based on the agreement of the fit for $\beta$, the two fits for $\varphi$, and the ODLRO behavior of Fig.~\ref{fig:PairingCorrelator}, we estimate that the critical point at $W/t=8$ is given by
\begin{equation}
    g_c=4.0514\pm 0.0005.
\end{equation}

The correlation-length exponent $\nu$ and the dynamic critical exponent $z$ can be obtained from the critical exponents determined above using relations between exponents. On the one hand, the exponent $\beta$ in Eq.~(\ref{beta}) is related to the scaling dimension $\Delta_{\partial_x\phi}$ of the PS order parameter field by the generalized hyperscaling relation $\beta=\nu\Delta_{\partial_x\phi}$~\cite{fisher1989}. Here $\Delta_{\partial_x\phi}=(d_\text{eff}-2+\eta)/2$ where $d_\text{eff}=d+z$ is the effective spacetime dimensionality ($d=1$ is the spatial dimension)~\cite{hertz1976}. On the other hand, the scaling dimension of the order parameter field obeys the exact relation $\Delta_{\partial_x\phi}=(3-z)/2$~\cite{nahum2025} which stems from the fact that the $\sim(\partial_\tau\phi)^2$ term in the effective action (\ref{Eq:Bosonized_S}) does not receive any perturbative corrections under renormalization. This is due to the fact that the interaction vertex $\sim(\partial_x\phi)^4$ only depends on the spatial derivative of $\phi$ and thus all perturbative corrections vanish at zero external momentum~\cite{andriyakhina2026}. Thus $\nu$ cancels in the ratio $\beta/\varphi=(3-z)/[2(z-1)]$, which can be solved for $z$ to give
\begin{align}
    z=\frac{2\beta+3\varphi}{2\beta+\varphi}=2.04\pm 0.05,
\end{align}
using the values quoted in Eqs.~(\ref{beta}) and (\ref{phi}). The correlation-length exponent $\nu$ is then found to be
\begin{align}
    \nu=0.61 \pm 0.05.
\end{align}
Those critical exponent values are close to the 2-loop field-theoretic predictions $z=1.99$ and $\nu=0.597$~\cite{flores-calderon2025} and to previous DMRG results~\cite{chepiga2026,shimomura2026}, but our inferred result for $z$ disagrees with the value $z\approx 1.5$ found in Ref.~\cite{flores-calderon2025}.

Finally, we turn to fermion single-particle properties at the critical point. In a conventional BCS superconductor with ODLRO, the equal-time electron Green's function $G(r)$, which is the Fourier transform of the momentum distribution $n(k)=\langle\hat{f}^\dag_k\hat{f}_k\rangle$,
\begin{align}\label{GF}
    G(r)\equiv\langle\hat{f}^\dag_0\hat{f}_r\rangle=\int\frac{dk}{2\pi}e^{ikr}n(k),
\end{align}
decays exponentially with a scale set by the coherence length. To see this, consider a 1D $p$-wave superconductor described by the BCS ground state $\ket{\Phi}=\prod_k\left(u_k+v_kc_k^\dag c_{-k}^\dag\right)\ket{0}$ with coherence factors $u_k,v_k$, i.e., the Kitaev chain~\cite{kitaev2001}, whose momentum distribution is
\begin{align}\label{nkBCS}
    n(k)=|v_k|^2=\frac{1}{2}\left(1-\frac{\epsilon_k}{\sqrt{\epsilon_k^2+|\Delta_k|^2}}\right),
\end{align}
at half filling ($\mu=0$), where $\epsilon_k$ is the single-particle dispersion and $\Delta_k=i\Delta\sin k$ is the $p$-wave gap function. The BCS momentum distribution (\ref{nkBCS}) is smooth and real-analytic. However, the long-distance behavior of $G(r)$ is controlled by the singularities of $n(k)$ in the complex plane that are closest to the real axis, which are the branch points at $k\pm k_F\approx\pm i\xi^{-1}$ where $\xi=v_F/\Delta$ is the superconducting coherence length. Expanding $n(k)$ near $\pm k_F$, we find for $r\gg a$,
\begin{align}\label{GBCS}
    G_\text{BCS}(r)\sim\sin(k_Fr)K_1(r/\xi)\sim\frac{\sin(k_Fr)}{\sqrt{|r|}}e^{-|r|/\xi},
\end{align}
where $K_1(z)$ is a modified Bessel function.

\begin{figure}[t]
\includegraphics[width=\columnwidth]{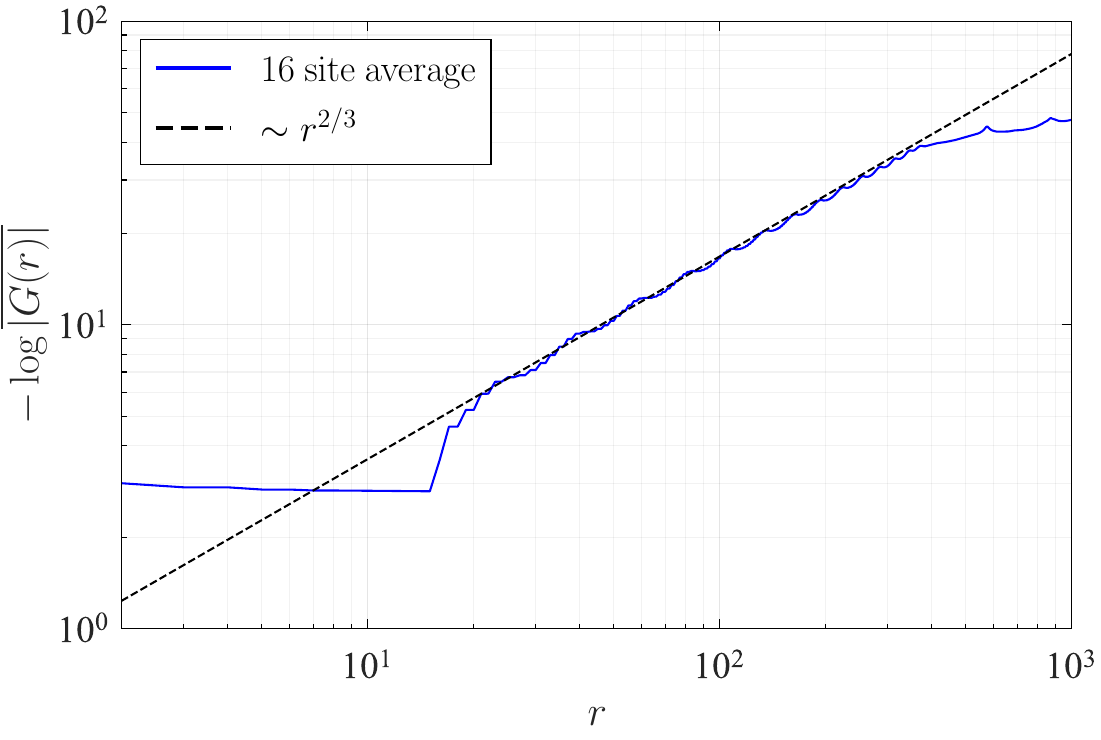}
\caption{Stretched-exponential behavior [Eq.~(\ref{StretchedExp})] of the fermionic Green's function $G(r)\equiv\langle\hat{f}^\dag_0\hat{f}_r\rangle$ computed from VUMPS at $g=g_c=4.0514$, for $W/t=8$ with a bond dimension $\chi=1600$. The absolute value $|G(r)|$ was averaged over a 16-site moving window to suppress $k_F$ oscillations. The dashed line is a manual fit to highlight the stretched-exponential behavior.}
\label{fig:stretchy}
\end{figure}

By contrast, we find using VUMPS in Fig.~\ref{fig:stretchy} that the fermionic Green's function (\ref{GF}) at the PS quantum critical point is not of the BCS exponential form (\ref{GBCS}) but rather shows \emph{stretched-exponential} decay at long distances,
\begin{align}\label{StretchedExp}
    G(r)\sim\frac{\sin(k_Fr)}{|r|^\sigma}e^{-C|r|^\alpha},\hspace{10mm}\alpha=\frac{2}{3},
\end{align}
a behavior first predicted using a large-$N$ field theory in Ref.~\cite{andriyakhina2026}. Here the subleading exponent $\sigma$ describes possible logarithmic corrections~\cite{andriyakhina2026} which we cannot accurately resolve in Fig.~\ref{fig:stretchy}. In Appendix~\ref{Sec:Critical_fermion_G}, we further show that the stretched-exponential behavior of the fermion Green's function (\ref{StretchedExp}) in real space implies that the momentum distribution $n(k)$ is smooth but \emph{nonanalytic} at $k=k_F$. This is thus a weak form of singular behavior that is intermediate between the smooth, real-analytic momentum distribution of BCS superconductors and the power-law singularity $n(k)-\half\propto\sgn(k-k_F)\cdot|k-k_F|^\delta$ in Luttinger liquids~\cite{giamarchi2003}. Away from PS criticality, the fermion Green's function is found to exhibit standard Luttinger-liquid behavior (see Appendix~\ref{sec:LLcheck}).

\section{Conclusion}
\label{sec:conclusion}

In summary, we have presented a number-conserving lattice model of locally interacting spinless fermions in 1D that exhibits long-range $p$-wave superconducting order at a zero-temperature QCP separating a Luttinger liquid from a phase-separated state. The mechanism for spontaneous breakdown of the $U(1)$ symmetry associated with particle number conservation is distinct from those requiring long-range interactions~\cite{maghrebi2017} or frustration-free Hamiltonians~\cite{watanabe2024}, but rather realizes a superconducting analog of a scenario recently proposed for XY ferromagnetism by Nahum~\cite{nahum2025}. What stabilizes off-diagonal long-range order in the present model is a quantum Berry phase term linking density and phase fluctuations in the critical low-energy theory, which invalidates the usual Mermin-Wagner obstruction to continuous symmetry breaking based on classical statistical mechanics. Based on a bosonized field-theory analysis and numerical methods including exact diagonalization and matrix-product state algorithms, our work provides a controlled, non-perturbative demonstration of a general idea explored earlier according to which critical density fluctuations in the vicinity of a phase-separation instability can enhance or even induce superconductivity in higher dimensions~\cite{grilli1991,perali1996,caprara2015}. Our best estimates for the critical exponents $\beta$ (order-parameter exponent), $\nu$ (correlation-length exponent), and $z$ (dynamic critical exponent) at the phase-separation transition are $\beta=0.29\pm 0.02$, $\nu=0.61\pm 0.05$, and $z=2.04\pm 0.05$, in good agreement with field-theoretic predictions and previous numerical studies~\cite{flores-calderon2025,chepiga2026,shimomura2026}. Additionally, we find numerical evidence, in agreement with Ref.~\cite{andriyakhina2026}, that the single-particle Green's function at criticality decays as a stretched exponential with stretching exponent $2/3$. We further show that this implies the single-particle momentum distribution $n(k)$ is a \emph{smooth nonanalytic function} at the Fermi wave vector $k=k_F$, a weak form of singularity intermediate between BCS superconductors and Luttinger liquids.

Our work opens up several avenues for future research. First, it would be interesting to construct locally interacting 1D models exhibiting long-range superconducting order with other pairing symmetries. One obvious direction to induce spin-singlet $s$-wave pairing in 1D would be to study XXZ-type Hamiltonians such as those considered in Ref.~\cite{nahum2025,flores-calderon2025,chepiga2026,shimomura2026,rey2026} but with $SU(2)$ spin operators replaced by $SU(2)$ pseudospin operators~\cite{zhang1990}, i.e., locally-interacting versions of the infinite-range Richardson model~\cite{richardson1963} supplemented with extended density-density interactions. It would also be interesting to study the effect of quenched disorder in the Hamiltonian (\ref{Hfermion}), since $p$-wave pairing is topologically robust in 1D~\cite{kitaev2001}, but randomness can change the universality class of the QCP~\cite{vojta2019} which, in the present scenario, induces pairing in the first place. Finally, it would be interesting to study the phenomenology of superconductivity in higher dimensions induced by critical density fluctuations in fermionic ladders or infinitely many coupled chains~\cite{bourbonnais1991,dagotto1992,schulz1996}.

\textit{Note added}: While this work was being independently finalized, Ref.~\cite{shimomura2026} appeared, which studies XY ferromagnetic long-range order in a spin-1/2 model equivalent via Jordan-Wigner transformation to our Hamiltonian (\ref{Hfermion}). Our study of an itinerant spinless fermion Hamiltonian provides a complementary perspective, including a derivation of the field theory from bosonization and a focus on distinct observables such as two-particle ODLRO ($p$-wave pairing) and the fermionic single-particle Green's function.

\acknowledgements

JM (FM) was supported by NSERC Discovery Grant No. RGPIN-2020-06999 (RGPIN-2022-03295). This research was enabled in part by support provided by the Digital Research Alliance of Canada (\href{https://www.alliancecan.ca}{alliancecan.ca}).

\onecolumngrid
\appendix
\numberwithin{equation}{section}
\numberwithin{figure}{section}

\section{Hartree-Fock ground-state energy}
\label{app:HF}

In this Appendix, we calculate the ground-state energy of the fermionic Hamiltonian (\ref{Hfermion}) in the Hartree-Fock approximation. The variational parameter is the PS order parameter $\OPS$ [Eq.~(\ref{OPS})], or equivalently, the density $n$ [Eq.~(\ref{n})]. The ground-state energy density $\varepsilon_\text{GS}(n)=\langle\hat{H}\rangle/L$ is
\begin{align}
\varepsilon_\text{GS}(n)=\varepsilon_0(n)+\varepsilon_\text{int}(n).
\end{align}
Here,
\begin{align}
\varepsilon_0(n)=\frac{1}{L}\langle\hat{H}_0\rangle=\frac{1}{L}\sum_k\varepsilon_k n_k,
\end{align}
is the energy of the free Fermi sea, where $n_k=\langle\n_k\rangle=\theta(\mu-\varepsilon_k)$ is the ground-state occupation number, and
\begin{align}
\varepsilon_\text{int}(n)=\frac{1}{L}\left(\langle\hat{H}_2\rangle+\langle\hat{H}_4\rangle\right),
\end{align}
is the energy due to interactions. The sum of the two interaction terms is
\begin{align}
\varepsilon_\text{int}(n)&=\frac{V}{4}+\frac{W}{16}-\left(V+\frac{W}{2}\right)n
+\frac{1}{L}\sum_r\left[\left(V+\frac{3W}{4}\right)\langle\hat{n}_r\hat{n}_{r+1}\rangle
+\frac{W}{2}\langle\hat{n}_r\hat{n}_{r+2}\rangle
+\frac{W}{4}\langle\hat{n}_r\hat{n}_{r+3}\rangle\right]\nn\\
&\phantom{=}-\frac{W}{2}\frac{1}{L}\sum_r
\left(2\langle\n_r\n_{r+1}\n_{r+2}\rangle+\langle\n_r\n_{r+1}\n_{r+3}\rangle+\langle\n_r\n_{r+2}\n_{r+3}\rangle\right)
+W\frac{1}{L}\sum_r\langle\n_r\n_{r+1}\n_{r+2}\n_{r+3}\rangle,
\end{align}
which contains two-body, three-body, and four-body terms that we evaluate in turn. We make use of the SNEG software package~\cite{zitko2011} to facilitate the evaluation of Wick contractions.

\subsection{Two-body terms}

\begin{figure}[h]
\begin{center}
\includegraphics[width=0.5\textwidth]{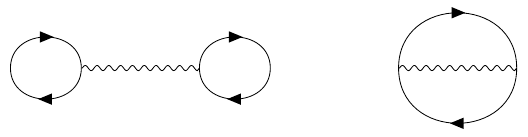}
\end{center}
\caption{Two-body Hartree (left) and Fock (right) contributions to the ground-state energy.}
\label{fig:HF-2body}
\end{figure}

We begin by normal ordering the interaction terms, which simplifies the evaluation of Wick contractions. Using the Fourier transform
\begin{align}
\f_r=\frac{1}{\sqrt{L}}\sum_k e^{ikr}\f_k,\hspace{5mm}\fd_r=\frac{1}{\sqrt{L}}\sum_k e^{-ikr}\fd_k,
\end{align}
and for any integer $\ell>0$, we have
\begin{align}\label{HF-2body}
\frac{1}{L}\sum_r\langle\n_r\n_{r+\ell}\rangle=\frac{1}{L}\sum_r\langle\fd_r\fd_{r+\ell}\f_{r+\ell}\f_r\rangle
=\frac{1}{L^2}\sum_{p_2p_3p_4}e^{i(-p_2+p_3)\ell}\langle\fd_{-p_2+p_3+p_4}\fd_{p_2}\f_{p_3}\f_{p_4}\rangle
=n^2-\chi(\ell),
\end{align}
which is nothing but the equal-time density-density correlation function of the free Fermi sea, with
\begin{align}
\chi(\ell)=\frac{1}{L^2}\sum_{kq}\cos(q\ell)n_kn_{k+q},
\end{align}
its connected part. The two terms in Eq.~(\ref{HF-2body}) correspond to the usual Hartree (direct) and Fock (exchange) terms, respectively (Fig.~\ref{fig:HF-2body}).

\subsection{Three-body terms}

\begin{figure}[h]
\begin{center}
\includegraphics[width=0.7\textwidth]{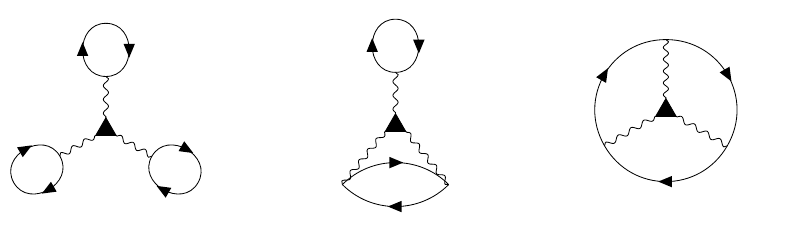}
\end{center}
\caption{Three-body interaction contributions to the ground-state energy.}
\label{fig:HF-3body}
\end{figure}

The three-body terms are all of the form $\langle\n_r\n_{r+\ell}\n_{r+m}\rangle$, with strictly positive integers $\ell\neq m$. We have
\begin{align}
\frac{1}{L}\sum_r\langle\n_r\n_{r+\ell}\n_{r+m}\rangle
=\frac{1}{L}\sum_r\langle\fd_r\fd_{r+\ell}\fd_{r+m}\f_{r+m}\f_{r+\ell}\f_r\rangle
=\frac{1}{L^3}\sum_{p_2\cdots p_6}e^{i(-p_2+p_5)\ell}e^{i(-p_3+p_4)m}
\langle\fd_{p_4+p_5+p_6-p_2-p_3}\fd_{p_2}\fd_{p_3}\f_{p_4}\f_{p_5}\f_{p_6}\rangle.
\end{align}
Upon performing the Wick contractions, we obtain the three-point density correlator
\begin{align}\label{HF-3body}
\frac{1}{L}\sum_r\langle\n_r\n_{r+\ell}\n_{r+m}\rangle=
n^3
-n\left[\chi(\ell)+\chi(m)+\chi(\ell-m)\right]
+\frac{1}{L^3}\sum_{kqq'}\left[\cos(q\ell+q'm)+\cos(qm+q'\ell)\right]n_kn_{k+q}n_{k+q+q'}.
\end{align}
The first term $n^3$ in Eq.~(\ref{HF-3body}) is a generalized Hartree diagram (Fig.~\ref{fig:HF-3body}, left); the second term can be viewed as a Fock diagram with a density-dependent ($\propto n$) effective two-body interaction (Fig.~\ref{fig:HF-3body}, middle); and the third term is a one-particle irreducible (1PI) diagram (Fig.~\ref{fig:HF-3body}, right) which can be viewed as a generalized, three-body Fock diagram. Note that the right-hand side of Eq.~(\ref{HF-3body}) is manifestly symmetric under interchange $\ell\leftrightarrow m$, as expected since the number operators $\n_{r+\ell}$ and $\n_{r+m}$ on the left-hand side commute.

\subsection{Four-body terms}

\begin{figure}[h]
\begin{center}
\includegraphics[width=0.9\textwidth]{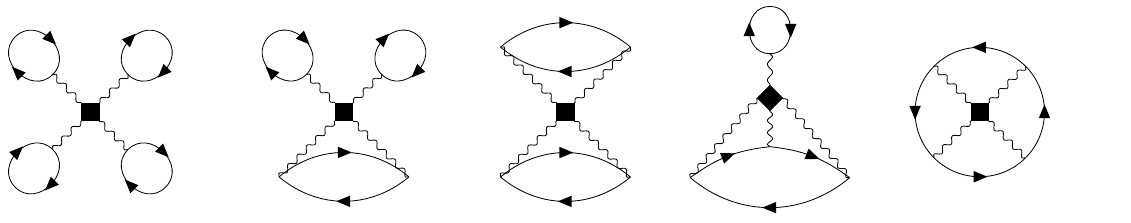}
\end{center}
\caption{Four-body interaction contributions to the ground-state energy.}
\label{fig:HF-4body}
\end{figure}

Finally, we consider the four-body term $\langle\n_r\n_{r+1}\n_{r+2}\n_{r+3}\rangle$. For the sake of generality, we evaluate $\langle\n_r\n_{r+\ell}\n_{r+m}\n_{r+s}\rangle$ with $\ell,m,s$ all distinct, strictly positive integers. We have
\begin{align}
\frac{1}{L}\sum_r\langle\n_r\n_{r+\ell}\n_{r+m}\n_{r+s}\rangle
&=\frac{1}{L}\sum_r\langle\fd_r\fd_{r+\ell}\fd_{r+m}\fd_{r+s}\f_{r+s}\f_{r+m}\f_{r+\ell}\f_r\rangle\nn\\
&=\frac{1}{L^4}\sum_{p_2\cdots p_8}e^{i(-p_2+p_7)\ell}e^{i(-p_3+p_6)m}e^{i(-p_4+p_5)s}\langle\fd_{p_5+p_6+p_7+p_8-p_2-p_3-p_4}\fd_{p_2}\fd_{p_3}\fd_{p_4}\f_{p_5}\f_{p_6}\f_{p_7}\f_{p_8}\rangle.
\end{align}
Performing the Wick contractions gives the four-point density correlator as
\begin{align}\label{HF-4body}
\frac{1}{L}\sum_r\langle\n_r\n_{r+\ell}\n_{r+m}\n_{r+s}\rangle&=
n^4
-n^2\left[\chi(\ell)+\chi(m)+\chi(s)+\chi(\ell-m)+\chi(m-s)+\chi(s-\ell)\right]\nn\\
&\phantom{=}+\chi(\ell-m)\chi(s)+\chi(m-s)\chi(\ell)+\chi(s-\ell)\chi(m)\nn\\
&\phantom{=}+\frac{n}{L^3}\sum_P\sum_{kqq'}\left\{\cos\left[qP(\ell)+q'P(m)\right]+\frac{1}{3}\cos\left[qP(\ell)+q'P(m)-(q+q')P(s)\right]\right\}n_kn_{k+q}n_{k+q+q'}\nn\\
&\phantom{=}-\frac{1}{L^4}\sum_P\sum_{kqq'q''}\cos\left[qP(\ell)+q'P(m)+q''P(s)\right]n_kn_{k+q}n_{k+q+q'}n_{k+q+q'+q''}.
\end{align}
The first term $n^4$ in Eq.~(\ref{HF-4body}) is a generalized four-body Hartree diagram (Fig.~\ref{fig:HF-4body}, leftmost diagram), and the term after that is a Fock diagram with a density-dependent ($\propto n^2$) effective two-body interaction (Fig.~\ref{fig:HF-4body}, second diagram from left). The second line of Eq.~(\ref{HF-4body}) is a product of two particle-hole bubbles (Fig.~\ref{fig:HF-4body}, third diagram from left). In the third and fourth lines of Eq.~(\ref{HF-4body}), the sum over $P$ denotes a sum over all $3!=6$ permutations of $\ell,m,s$. As a result, the entire right-hand side of Eq.~(\ref{HF-4body}) is symmetric under permutations of $\ell,m,s$, which is expected since the number operators $\n_{r+\ell}$, $\n_{r+m}$, $\n_{r+s}$ on the left-hand side mutually commute. The third line of Eq.~(\ref{HF-4body}), corresponding to the fourth diagram from the left in Fig.~\ref{fig:HF-4body}, can be viewed as the rightmost diagram of Fig.~\ref{fig:HF-3body} but with a density-dependent ($\propto n$) three-body interaction vertex. Finally, the fourth line of Eq.~(\ref{HF-4body}) corresponds to the rightmost diagram of Fig.~\ref{fig:HF-4body}.

\section{Bosonization details} \label{Sec:Bosonization}

This Appendix provides details about the bosonization derivations and calculations. The lattice spacing $a$ is reintroduced for the first two sections.

\subsection{Derivation of low-energy theory}\label{Sec:Bosonization_H}

This first section presents the derivation of the low-energy field theory describing the phase-separation quantum phase transition. The theory is obtained by approaching the critical point from the disordered phase using Abelian bosonization. 

We start from the 1D lattice Hamiltonian (\ref{Hfermion}),
\begin{align}
\begin{split}
\hat{H} = -t \sum_r \Big( \hat{f}_r^{\dag} \hat{f}_{r+1} + \text{H.c.} \Big) + V \sum_r \delta \hat{n}_r \, \delta \hat{n}_{r+1} + W \sum_r \delta \hat{n}_r \, \delta \hat{n}_{r+1} \, \delta \hat{n}_{r+2} \, \delta \hat{n}_{r+3} \equiv \hat{H}_0 + \hat{H}_2 + \hat{H}_4 ,
\end{split}
\end{align}
where $\delta \hat{n}_r = \hat{f}_r^{\dag} \hat{f}_r -1/2$ is the density measured with respect to half filling. The bosonization procedure starts by expanding the lattice fermion operator near the two Fermi points:
\begin{equation}\label{Eq:Exp_lattice_fermion}
\hat{f}_r \approx \sqrt{a} \Big( e^{i k_F x} \psi_R(x) + e^{-i k_F x} \psi_L(x) \Big) ,
\end{equation}
where $\psi_{R/L}(x)$ is a slowly-varying right/left-moving fermion operator, $x= r a$, where $a$ is the lattice spacing, and $k_F$ is the Fermi momentum, which at half filling is $k_F = \frac{\pi}{2a}$. The bosonized form of the slowly-varying fields is \cite{giamarchi2003}
\begin{equation}\label{Eq:psi_eta}
\psi_{\eta}(x) = \frac{U_{\eta}}{\sqrt{2\pi \alpha}} e^{-i [\eta \phi(x) - \theta(x)]} ,
\end{equation}
where $\alpha$ is a UV cutoff, $\eta = R/L$ $(\pm 1)$ and $U_{\eta}$ are the Klein factors ensuring the fermionic anticommutation relations, which respect $\{U_{\eta},U_{\eta'}\} = \{U_{\eta},U_{\eta'}^{\dag}\} = 0$ for $\eta\neq \eta'$ and $U_{\eta} U_{\eta}^{\dag} = U_{\eta}^{\dag} U_{\eta} = 1$. The two periodic scalar fields $\phi(x)$ and $\theta(x)$ satisfy
\begin{equation}\label{CanonicalCommutators}
[\phi(x),\theta(x')] = i \frac{\pi}{2} \sgn(x'-x) , \qquad [\phi(x),\frac{1}{\pi} \partial_{x'} \theta(x')] = i \delta(x'-x) ,
\end{equation}
where the sign function is defined such that $\sgn(0)=0$. The second commutation relation implies that $\frac{1}{\pi} \partial_x\theta(x) = \Pi(x)$ is the canonical momentum conjugate to $\phi(x)$. Using this, we can proceed with the bosonization of $\hat{H}$.

\subsubsection{Bosonization of $\hat{H}_0$}

We start with the free Hamiltonian $\hat{H}_0$. Using Eq. (\ref{Eq:Exp_lattice_fermion}), we get
\begin{align}
\begin{split}
\hat{H}_0 &= -t a \sum_r \Big[ e^{i k_F a} \psi_R^{\dag}(x) \psi_R(x+a) + e^{-i k_F a} \psi_L^{\dag}(x) \psi_L(x+a) \\ &\hspace{2cm} + e^{-i 2 k_F x} e^{-i k_F a} \psi_R^{\dag}(x) \psi_L(x+a) + e^{i 2 k_F x} e^{i k_F a} \psi_L^{\dag}(x) \psi_R(x+a) +\text{H.c.} \Big] \\ &\approx -t \int dx \Big[ e^{i k_F a} \psi_R^{\dag}(x) \psi_R(x+a) + e^{-i k_F a} \psi_L^{\dag}(x) \psi_L(x+a) + \text{H.c.} \Big] ,
\end{split}
\end{align}
where the sum over lattice sites was converted to a spatial integral in the second equality using $\sum_r \rightarrow \frac{1}{a} \int dx$. The third and fourth terms in the first equality correspond to the rapidly oscillating $q \sim 2k_F$ contributions, which when integrated over a slowly-varying piece, average out to zero. The product of chiral fermion operators is now bosonized using Eq. (\ref{Eq:psi_eta}). We first have
\begin{align}
\begin{split}
\psi_R^{\dag}(x) \psi_R(x+a) = \frac{U_R^{\dag} U_R}{2\pi \alpha} e^{i[\phi(x)-\theta(x)]} e^{-i[\phi(x+a)-\theta(x+a)]} = \frac{-i}{2\pi \alpha} e^{-i \Delta\phi(x) + i \Delta\theta(x)} ,
\end{split}
\end{align}
where we have defined $\Delta\phi(x) = \phi(x+a)-\phi(x)$ and $\Delta \theta(x) = \theta(x+a)-\theta(x)$. The two exponentials have been combined by noting that
\begin{equation}
[i(\phi(x)-\theta(x)),(-i)(\phi(x+a)-\theta(x+a))] = -i\frac{\pi}{2} \sgn(a) + i \frac{\pi}{2} \sgn(-a) = -i \pi ,
\end{equation}
and by using the truncated Baker-Campbell-Hausdorff (BCH) formula $e^A e^B = e^{A+B} e^{\frac{1}{2}[A,B]}$. Similarly, one can show that
\begin{equation}
\psi_L^{\dag}(x) \psi_L(x+a) = \frac{i}{2\pi \alpha} e^{i \Delta\phi(x) + i \Delta\theta(x)} .
\end{equation}
Hence, the free Hamiltonian becomes
\begin{align}
\begin{split}
\hat{H}_0 &\approx - \frac{t}{2\pi \alpha} \int dx \Big[ e^{-i \Delta\phi(x) + i \Delta\theta(x)} + e^{i \Delta\phi(x) + i \Delta\theta(x)} + e^{i \Delta\phi(x) - i \Delta\theta(x)} + e^{-i \Delta\phi(x) - i \Delta\theta(x)} \Big] \\ &= - \frac{t}{2\pi \alpha} 2 \int dx \Big[ \cos\big( \Delta\phi(x) - \Delta\theta(x) \big) + \cos\big( \Delta\phi(x) + \Delta\theta(x) \big) \Big] ,
\end{split}
\end{align}
where we have used that $e^{\pm i k_F a} = \pm i$ at half filling. Furthermore, the exponentials could be combined into cosines since $[\Delta\phi(x),\Delta\theta(x)]=0$. These are then expanded to fourth order, yielding
\begin{align}
\begin{split}
\hat{H}_0 &\approx - \frac{t}{2\pi \alpha} 2 \int dx \Big[ - (\Delta\phi(x))^2 - (\Delta\theta(x))^2 + \frac{1}{12} \Big( (\Delta\phi(x))^4 + (\Delta\theta(x))^4 \Big) + \frac{1}{2} (\Delta\phi(x))^2 (\Delta\theta(x))^2 \Big] ,
\end{split}
\end{align}
followed by a fourth-order gradient expansion. For the first term, we have
\begin{equation}
(\Delta\phi(x))^2 \approx a^2 (\partial_x\phi)^2 + a^3 \partial_x\phi \partial_x^2\phi + \frac{a^4}{4} (\partial_x^2\phi)^2 + \frac{a^4}{3}\partial_x\phi \partial_x^3\phi + \dots \approx a^2 (\partial_x\phi)^2 - \frac{a^4}{12}(\partial_x^2\phi)^2 + \dots ,
\end{equation}
where the second term in the first equality was dropped since it is a total derivative, while the fourth term was integrated by parts. These two manipulations will be used extensively in the rest of this Appendix. Expanding the other terms in a similar manner yields
\begin{align}
\begin{split}
(\Delta \theta(x))^2 &\approx a^2 (\partial_x\theta)^2 - \frac{a^4}{12} (\partial_x^2\theta)^2 + \dots \\ (\Delta \phi(x))^4 &\approx a^4 (\partial_x\phi)^4 + \dots \\ (\Delta \theta(x))^4 &\approx a^4 (\partial_x\theta)^4 + \dots \\ (\Delta\phi(x))^2 (\Delta \theta(x))^2 &\approx a^4 (\partial_x\phi(x))^2 (\partial_x\theta(x))^2 + \dots . 
\end{split}
\end{align}
At quartic order in derivatives, the bosonized version of the free Hamiltonian is thus
\begin{align}\label{Eq:Bosonized_H0}
\begin{split}
\hat{H}_0 &= \int dx \Bigg( \frac{ta}{\pi} \Big[ (\partial_x\phi)^2 + (\partial_x\theta)^2 \Big] - \frac{a^3t}{12\pi} \Big[ (\partial_x^2\phi)^2 + (\partial_x^2\theta)^2 \Big] \\ &\hspace{1.5cm} - \frac{a^3t}{12\pi} \Big[ (\partial_x\phi)^4 + (\partial_x\theta)^4 \Big] - \frac{a^3 t}{4\pi} \Big[ (\partial_x\phi)^2 (\partial_x\theta)^2 + (\partial_x\theta)^2 (\partial_x\phi)^2 \Big] \Bigg) + \dots ,
\end{split}
\end{align}
where the UV cutoff $\alpha$ was set equal to the lattice spacing $a$.

\subsubsection{Bosonization of $\hat{H}_2$}

The bosonization of the two-body interaction first requires to bosonize the density operator. Using Eqs. (\ref{Eq:Exp_lattice_fermion}) and (\ref{Eq:psi_eta}) and following Ref.~\cite{giamarchi2003}, we get
\begin{align}
\begin{split}
\hat{f}_r^{\dag} \hat{f}_r &= a \Big[ \psi_R^{\dag}(x) \psi_R(x) + \psi_L^{\dag}(x) \psi_L(x) + e^{-i 2 k_F x} \psi_R^{\dag}(x) \psi_L(x) + e^{i 2 k_F x} \psi_L^{\dag}(x) \psi_R(x) \Big] \\ &\approx a \Bigg[ \rho_0 - \frac{1}{\pi} \partial_x\phi(x) + e^{-i 2 k_F x} \frac{U_R^{\dag} U_L}{2\pi \alpha} e^{i[\phi(x) - \theta(x)]} e^{i[\phi(x) + \theta(x)]} + e^{i 2 k_F x} \frac{U_L^{\dag} U_R}{2\pi \alpha} e^{-i[\phi(x) + \theta(x)]} e^{-i[\phi(x) - \theta(x)]} \Bigg] + \dots ,
\end{split}
\end{align}
where $\rho_0 = \frac{1}{2a}$ is the density at half filling, while we also have $e^{\pm i 2 k_F x} = e^{\pm i \pi r} = (-1)^r$. Furthermore, since $[\phi(x),\theta(x)]=0$, the product of exponentials can be combined. This results in the following expression for the lattice density operator
\begin{equation}\label{Eq:Bosonized_density}
\delta \hat{n}_r \approx a \Bigg[ -\frac{1}{\pi} \partial_x\phi(x) + \frac{(-1)^r}{\pi \alpha} \cos(2\phi(x)) \Bigg] + \dots ,
\end{equation}
where the Klein factors will be omitted from now on \cite{giamarchi2003}. 

Injecting the above expression in the two-body Hamiltonian yields
\begin{align}
\begin{split}
\hat{H}_2 &\approx V a^2 \sum_r \Bigg[ \frac{1}{\pi^2} \partial_x\phi(x) \partial_x\phi(x+a) - \frac{1}{(\pi \alpha)^2} \cos(2\phi(x)) \cos(2\phi(x+a)) \\ &\hspace{2cm} + \frac{(-1)^r}{\pi^2 \alpha} \Big( \partial_x\phi(x) \cos(2\phi(x+a)) - \partial_x\phi(x+a) \cos(2\phi(x)) \Big) \Bigg] + \dots \\ &\approx V a \int dx \Bigg[ \frac{1}{\pi^2} \partial_x\phi(x) \partial_x\phi(x+a) - \frac{1}{2(\pi \alpha)^2} \Big( \cos\big( 2\phi(x)-2\phi(x+a) \big) + \cos\big( 2\phi(x)+2\phi(x+a) \big) \Big) \Bigg] + \dots ,
\end{split}
\end{align}
where the $q \sim 2k_F$ terms in the first equality average out to zero, while the trigonometric identity $\cos \vartheta \, \cos \varphi = \frac{1}{2} \big( \cos(\vartheta-\varphi) + \cos(\vartheta-\varphi) \big)$ was used to rewrite the product of  cosines in the first line. Expanding the two terms to fourth order in the gradient expansion yields
\begin{align}
\begin{split}
\hat{H}_2 &\approx V a \int dx \Bigg[ \frac{1}{\pi^2} \Bigg( (\partial_x\phi)^2 + a \partial_x\phi \partial_x^2\phi + \frac{a^2}{2} \partial_x\phi \partial_x^3\phi \Bigg) - \frac{1}{2(\pi \alpha)^2} \Bigg( 1 - 2 a^2 (\partial_x\phi)^2 \\ &\hspace{2cm} - 2 a^3 \partial_x\phi \partial_x^2\phi + \frac{2}{3} a^4 (\partial_x\phi)^4 - \frac{a^4}{2} (\partial_x^2\phi)^2 - \frac{2}{3}a^4 \partial_x\phi \partial_x^3\phi + \cos(4\phi) + \text{Mixed terms} \Bigg)\Bigg] + \dots ,
\end{split}
\end{align}
where ``Mixed terms'' stands for terms containing derivatives of $\phi(x)$ as well as either $\cos(4\phi(x))$ or $\sin(4\phi(x))$, which are of higher order and can be neglected. After setting $\alpha = a$ and some simplifications, we finally get
\begin{align}
\begin{split}
\hat{H}_2 \approx V a \int dx \Bigg[ \frac{2}{\pi^2} (\partial_x\phi)^2 - \frac{7}{12\pi^2} a^2 (\partial_x^2\phi)^2 - \frac{a^2}{3\pi^2} (\partial_x\phi)^4 - \frac{1}{2(\pi \alpha)^2} \cos(4\phi) \Bigg] + \dots .
\end{split}
\end{align}

\subsubsection{Bosonization of $\hat{H}_4$}

We finally move to the bosonization of the four-body interaction. Using Eq. (\ref{Eq:Bosonized_density}), we get

\begin{align}
\begin{split}
\hat{H}_4 &\approx W a^3 \int dx \Bigg[ \frac{1}{\pi^4} \partial_x\phi(x) \partial_x\phi(x+a) \partial_x\phi(x+2a) \partial_x\phi(x+3a) \\ &\hspace{2cm} - \frac{1}{\pi^4 \alpha^2} \partial_x\phi(x) \partial_x\phi(x+a) \cos(2\phi(x+2a)) \cos(2\phi(x+3a)) \\ &\hspace{2cm} + \frac{1}{\pi^4 \alpha^2} \partial_x\phi(x) \partial_x\phi(x+2a) \cos(2\phi(x+a)) \cos(2\phi(x+3a)) \\ &\hspace{2cm} - \frac{1}{\pi^4 \alpha^2} \partial_x\phi(x) \partial_x\phi(x+3a) \cos(2\phi(x+a)) \cos(2\phi(x+2a)) \\ &\hspace{2cm} - \frac{1}{\pi^4 \alpha^2} \partial_x\phi(x+a) \partial_x\phi(x+2a) \cos(2\phi(x)) \cos(2\phi(x+3a)) \\ &\hspace{2cm} + \frac{1}{\pi^4 \alpha^2} \partial_x\phi(x+a) \partial_x\phi(x+3a) \cos(2\phi(x)) \cos(2\phi(x+2a)) \\ &\hspace{2cm} - \frac{1}{\pi^4 \alpha^2} \partial_x\phi(x+2a) \partial_x\phi(x+3a) \cos(2\phi(x)) \cos(2\phi(x+a)) \\ &\hspace{2cm} + \frac{1}{(\pi \alpha)^4} \cos(2\phi(x)) \cos(2\phi(x+a)) \cos(2\phi(x+2a)) \cos(2\phi(x+3a)) \Bigg] \\ &\equiv a^3 W \int dx \Big[ T_1 + T_2 + T_3 + T_4 + T_5 + T_6 + T_7 + T_8 \Big] ,
\end{split}
\end{align}
where the eight $q \sim 2k_F$ terms average out to zero as before. Once again, we expand this expression to fourth order in derivatives. For $T_1$, we simply have
\begin{equation}\label{Eq:T_1}
T_1 = \frac{1}{\pi^4} \partial_x\phi(x) \partial_x\phi(x+a) \partial_x\phi(x+2a) \partial_x\phi(x+3a) \approx \frac{1}{\pi^4} (\partial_x\phi)^4 + \dots .
\end{equation}

The expansion of $T_2$ is as follows
\begin{align}\label{Eq:T_2}
\begin{split}
T_2 &= - \frac{1}{\pi^4 \alpha^2} \partial_x\phi(x) \partial_x\phi(x+a) \cos(2\phi(x+2a)) \cos(2\phi(x+3a)) \\ &= - \frac{1}{2\pi^4 \alpha^2} \partial_x\phi(x) \partial_x\phi(x+a) \Big[ \cos\big( 2\phi(x+2a) - 2\phi(x+3a) \big) + \cos\big( 2\phi(x+2a) + 2\phi(x+3a) \big) \Big] \\ &\approx -\frac{1}{2\pi^4 \alpha^2} \Bigg[ (\partial_x\phi)^2 + a \partial_x\phi \partial_x^2\phi -2 a^2 (\partial_x\phi)^4  + \frac{a^2}{2} \partial_x\phi \partial_x^3\phi + \text{Mixed terms} \Bigg] \\ &\approx -\frac{1}{2\pi^4 \alpha^2} \Bigg[ (\partial_x\phi)^2 -2 a^2 (\partial_x\phi)^4  - \frac{a^2}{2} (\partial_x^2\phi)^2 \Bigg] + \dots ,
\end{split}
\end{align}
where as before, a total derivative was dropped, integration by parts was used, and terms mixing derivatives of $\phi$ and either $\cos(4\phi(x))$ or $\sin(4\phi(x))$ were dropped. Proceeding similarly for $T_3$ to $T_7$, we get
\begin{align}\label{Eq:T_3-T_7}
\begin{split}
T_3 &\approx \frac{1}{2\pi^4 \alpha^2} \Bigg[ (\partial_x\phi)^2 - 8 a^2 (\partial_x\phi)^4 - 2 a^2 (\partial_x^2\phi)^2 \Bigg] + \dots \, , \quad T_4 \approx -\frac{1}{2\pi^4 \alpha^2} \Bigg[ (\partial_x\phi)^2 - 2 a^2 (\partial_x\phi)^4 - \frac{9}{2} a^2 (\partial_x^2\phi)^2 \Bigg] + \dots \, , \\ T_5 &\approx -\frac{1}{2\pi^4 \alpha^2} \Bigg[ (\partial_x\phi)^2 -18 a^2 (\partial_x\phi)^4 - \frac{a^2}{2} (\partial_x^2\phi)^2 \Bigg] + \dots \, , \quad T_6 \approx \frac{1}{2\pi^4 \alpha^2} \Bigg[ (\partial_x\phi)^2 -8 a^2 (\partial_x\phi)^4 - 2 a^2 (\partial_x^2\phi)^2 \Bigg] + \dots \, , \\ T_7 &\approx -\frac{1}{2\pi^4 \alpha^2} \Bigg[ (\partial_x\phi)^2 - 2 a^2 (\partial_x\phi)^4 - \frac{a^2}{2} (\partial_x^2\phi)^2 \Bigg] + \dots .
\end{split}
\end{align}

Finally, the expansion of $T_8$ is the most cumbersome. The product of four cosines is first written as
\begin{align}
\begin{split}
T_8 &= \frac{1}{(\pi \alpha)^4} \cos(2\phi(x)) \cos(2\phi(x+a)) \cos(2\phi(x+2a)) \cos(2\phi(x+3a)) 
\\ &= \frac{1}{4(\pi\alpha)^4} \Big[ \cos\big( 2\phi(x) - 2\phi(x+a) \big) + \cos\big( 2\phi(x) + 2\phi(x+a) \big) \Big] \\ &\hspace{1.3cm}\times \Big[ \cos\big( 2\phi(x+2a) - 2\phi(x+3a) \big) + \cos\big( 2\phi(x+2a) + 2\phi(x+3a) \big) \Big] 
\\ &= \frac{1}{8(\pi\alpha)^4} \Big[ \cos\big( 2\phi(x)-2\phi(x+a)-2\phi(x+2a) + 2\phi(x+3a) \big) \\ &\hspace{2cm}+ \cos\big( 2\phi(x)-2\phi(x+a)+2\phi(x+2a) - 2\phi(x+3a) \big) \\ &\hspace{2cm}+ \cos\big( 2\phi(x)-2\phi(x+a)-2\phi(x+2a) - 2\phi(x+3a) \big) \\ &\hspace{2cm}+ \cos\big( 2\phi(x)-2\phi(x+a)+2\phi(x+2a) + 2\phi(x+3a) \big) \\ &\hspace{2cm}+ \cos\big( 2\phi(x)+2\phi(x+a)-2\phi(x+2a) + 2\phi(x+3a) \big) \\ &\hspace{2cm}+ \cos\big( 2\phi(x)+2\phi(x+a)+2\phi(x+2a) - 2\phi(x+3a) \big) \\ &\hspace{2cm}+ \cos\big( 2\phi(x)+2\phi(x+a)-2\phi(x+2a) - 2\phi(x+3a) \big) \\ &\hspace{2cm}+ \cos\big( 2\phi(x)+2\phi(x+a)+2\phi(x+2a) + 2\phi(x+3a) \big) \Big] .
\end{split}
\end{align}
Expanding the eight terms in powers of $a$ and simplifying results in
\begin{align}\label{Eq:T_8}
\begin{split}
T_8 &\approx \frac{1}{8(\pi \alpha)^4} \Bigg[ \Big( 1 - 8 a^4 (\partial_x^2\phi)^2 +\dots \Big) \\ &\hspace{1.3cm}+ \Big( 1 - 8 a^2 (\partial_x\phi)^2 - 24 a^3 \partial_x\phi \partial_x^2\phi + \frac{32}{3}a^4 (\partial_x\phi)^4 - \frac{54}{3} a^4 (\partial_x^2\phi)^2 - \frac{80}{3} a^4 \partial_x\phi \partial_x^3\phi + \dots \Big) \\ &\hspace{1.3cm} + 4 \Big( \cos(4\phi) + \text{Mixed terms} \Big) \\ &\hspace{1.3cm}+ \Big( 1 - 32 a^2 (\partial_x\phi)^2 - 96 a^3 \partial_x\phi \partial_x^2\phi + \frac{512}{3}a^4 (\partial_x\phi)^4 - 72 a^4 (\partial_x^2\phi)^2 - \frac{272}{3} a^4 \partial_x\phi \partial_x^3\phi + \dots \Big) \\ &\hspace{1.3cm} + \Big( \cos(8\phi) + \text{Mixed terms} \Big) \Bigg] 
\\ &\approx \frac{1}{8(\pi \alpha)^4} \Bigg[ \Big( 1 - 8 a^4 (\partial_x^2\phi)^2 \Big) + 4 \cos(4\phi) + \Big( 1 - 8 a^2 (\partial_x\phi)^2 + \frac{32}{3}a^4 (\partial_x\phi)^4 + \frac{26}{3} a^4 (\partial_x^2\phi)^2 \Big) \\ &\hspace{1.3cm}+ \Big( 1 - 32 a^2 (\partial_x\phi)^2 + \frac{512}{3}a^4 (\partial_x\phi)^4 +\frac{56}{3} a^4 (\partial_x^2\phi)^2 \Big) \Bigg] + \dots \\ &\approx \frac{1}{8(\pi\alpha)^4} \Bigg[ - 40 a^2 (\partial_x\phi)^2 + \frac{544}{3} a^4 (\partial_x\phi)^4 + \frac{58}{3} a^4 (\partial_x^2\phi)^2 + 4 \cos(4\phi) \Bigg] + \dots .
\end{split}
\end{align}

By combining the eight terms of Eqs. (\ref{Eq:T_1}-\ref{Eq:T_8}) and simplifying, we finally get the following expression for the bosonized four-body interaction
\begin{align}\label{Eq:Bosonized_H4}
\begin{split}
\hat{H}_4 \approx a^3 W \int dx \Bigg[ -\frac{6}{\pi^4 a^2} (\partial_x\phi)^2 + \frac{41}{12\pi^4} (\partial_x^2\phi)^2 + \frac{83}{3\pi^4} (\partial_x\phi)^4 + \frac{1}{2 \pi^4 a^4} \cos(4\phi) \Bigg] ,
\end{split}
\end{align}
where once again, the UV cutoff $\alpha$ was replaced by the lattice spacing $a$.

\subsubsection{Low-energy Euclidean action}

Using the results of the three previous sections, we get the full expression for the bosonized Hamiltonian:
\begin{align}\label{Eq:Bosonized_H}
\begin{split}
\hat{H} &= \int dx \Bigg[ \frac{a t}{\pi} (\partial_x\theta)^2 + \frac{a}{\pi} \Big( t + \frac{2}{\pi} V - \frac{6}{\pi^3} W \Big) (\partial_x\phi)^2 + \frac{a^3}{12\pi} \Big(-t -\frac{7}{\pi}V + \frac{41}{\pi^3} W \Big) (\partial_x^2\phi)^2 \\ &\hspace{1.5cm} + \frac{a^3}{12\pi} \Big(-t -\frac{4}{\pi}V + \frac{332}{\pi^3} W \Big) (\partial_x\phi)^4 + \frac{1}{2\pi^2 a} \Big( -V + \frac{1}{\pi^2} W \Big) \cos(4\phi) \\ &\hspace{1.5cm} - \frac{t a^3}{12\pi} (\partial_x^2\theta)^2 - \frac{t a^3}{12\pi} (\partial_x\theta)^4 - \frac{t a^3}{2\pi} (\partial_x \phi)^2 (\partial_x\theta)^2 \Bigg] + \dots 
\\ &= \int dx \Big[ c_1 (\pi \Pi)^2 + c_2 (\partial_x\phi)^2 + c_3 (\partial_x^2\phi)^2 + c_4 (\partial_x\phi)^4 + c_5 \cos(4\phi) \\ &\hspace{1.5cm} - c_6 (\pi\partial_x\Pi)^2 - c_7 (\pi \Pi)^4 - c_8 (\partial_x \phi)^2 (\pi \Pi)^2 \Big] + \dots ,
\end{split}
\end{align}
where $c_1 , ..., c_8$ are the coefficients in front of the eight terms, while we also recall that $\partial_x\theta(x) = \pi \Pi(x)$.

The imaginary-time action is obtained by performing the inverse Legendre transform:
\begin{align}\label{Eq:S_v1}
\begin{split}
S[\phi,\theta] &= -i \int d\tau \, dx\, \Pi \partial_{\tau}\phi + \int d\tau \, H \\ &= \int d\tau \, dx \Big[ -\frac{i}{\pi} \partial_x\theta \partial_{\tau}\phi + c_1 (\partial_x\theta)^2 + c_2 (\partial_x\phi)^2 + c_3 (\partial_x^2\phi)^2 + c4 (\partial_x\phi)^4 + c_5 \cos(4\phi) \\ &\hspace{2.5cm} - c_6 (\partial_x^2\theta)^2 - c_7 (\partial_x\theta)^4 - c_8 (\partial_x\phi)^2 (\partial_x\theta)^2 \Big] + \dots .
\end{split}
\end{align}
The effective action $S[\phi]$ is obtained by integrating out the smooth Gaussian fluctuations of $\theta$. We thus neglect the quartic interaction $c_7 (\partial_x\theta)^4$, whose inclusion would further renormalize the coefficients of the leading operators in the low-energy theory, as well as generate additional higher-order irrelevant operators. By defining $\psi = \partial_x\theta$, all the terms containing $\theta$ are grouped into
\begin{align}
\begin{split}
S_{\theta}[\phi,\theta] = \int d\tau \, dx \, \Big[ -\frac{i}{\pi} \psi \partial_{\tau} \phi + c_1 \psi^2 -c_6 (\partial_x\psi)^2 - c_8 (\partial_x\phi)^2 \psi^2 \Big] \equiv \int d\tau \, dx \, \Big[ -\frac{i}{\pi} \psi \partial_{\tau} \phi + \psi \hat{O}(\phi) \psi \Big] ,
\end{split}
\end{align}
where the operator $\hat{O}(\phi) = c_1 + c_6 \partial_x^2 - c_8 (\partial_x\phi)^2$ was introduced. The path integral over $\psi$ is thus performed by completing the square and using the change of variable $\Tilde{\psi} = \psi - \frac{i}{2\pi} \hat{O}^{-1}(\phi) \dot{\phi}$, where $\dot{\phi} = \partial_{\tau} \phi$:
\begin{align}
\begin{split}
\int \mathcal{D}[\psi] e^{-S_{\theta}[\phi,\theta]} &= \exp\Bigg( -\frac{1}{4\pi^2} \int d\tau dx \, \dot{\phi} \hat{O}^{-1}(\phi) \dot{\phi} \Bigg) \int \mathcal{D}[\Tilde{\psi}] \exp\Bigg( -\int d\tau \, dx \, \Tilde{\psi} \hat{O}(\phi) \Tilde{\psi} \Bigg) \\ &= \mathcal{N} \exp\Bigg( -\frac{1}{2} \Tr \log \hat{O}(\phi) \Bigg) \exp\Bigg( -\frac{1}{4\pi^2} \int d\tau dx \, \dot{\phi} \hat{O}^{-1}(\phi) \dot{\phi} \Bigg) ,
\end{split}
\end{align}
where $\mathcal{N}$ is an unimportant normalization constant. The full effective action for $\phi$ is thus
\begin{align}\label{Eq:S_eff_phi}
\begin{split}
S[\phi] &= \frac{1}{4\pi^2} \int d\tau dx \, \dot{\phi} \hat{O}^{-1}(\phi) \dot{\phi} + \frac{1}{2} \Tr \log \hat{O}(\phi) \\ &\hspace{0.7cm} + \int d\tau dx \, \Big[ c_2 (\partial_x\phi)^2 + c_3 (\partial_x^2\phi)^2 + c_4 (\partial_x\phi)^4 + c_5 \cos(4\phi) \Big] + \dots .
\end{split}
\end{align}

The first two terms of the above equation must now be expanded. We have for the first one
\begin{align}
\begin{split}
\frac{1}{4\pi^2} \int d\tau dx \, \dot{\phi} \hat{O}^{-1}(\phi) \dot{\phi}  &= \frac{1}{4\pi^2 c_1} \int d\tau \, dx \, \dot{\phi} \Big( 1 + \frac{c_6}{c_1} \partial_x^2 - \frac{c_8}{c_1} (\partial_x\phi)^2 \Big)^{-1} \dot{\phi} \\ &\approx \frac{1}{4\pi^2 c_1} \int d\tau \, dx \, \dot{\phi} \Big( 1 - \frac{c_6}{c_1} \partial_x^2 + \frac{c_8}{c_1} (\partial_x\phi)^2 \Big) \dot{\phi} + \dots \\ &\approx \frac{1}{4\pi^2 c_1} \int d\tau \, dx \, (\partial_{\tau}\phi)^2 + \dots ,
\end{split}
\end{align}
where the higher-order operators $\dot{\phi} \partial_x^2 \dot{\phi}$ and $\dot{\phi}^2 (\partial_x\phi)^2$ are strongly irrelevant and can safely be dropped.

We now move on to the expansion of the second term of Eq. (\ref{Eq:S_eff_phi}). To do so, we first define the inverse free Green's function $G_0^{-1} = c_1 + c_6 \partial_x^2$. Hence
\begin{align}
\begin{split}
\frac{1}{2} \Tr \log \hat{O}(\phi) &= \frac{1}{2} \Tr \log \Big[ G_0^{-1} \Big( 1 - c_8 G_0 (\partial_x\phi)^2 \Big) \Big] \\ &\approx \frac{1}{2} \Tr \log G_0^{-1} - \frac{c_8}{2} \Tr \Big[ G_0 (\partial_x\phi)^2 \Big] - \frac{c_8^2}{4} \Tr \Big[ G_0 (\partial_x\phi)^2 G_0 (\partial_x\phi)^2 \Big] + \dots .
\end{split}
\end{align}
The first term is a constant and will be dropped. To evaluate the next two terms, we will use the following expression for the free Green's function:
\begin{equation}
G_0(\tau,x) = \int \frac{d\omega\,dk}{(2\pi)^2} e^{i \omega \tau} e^{i k x} \Tilde{G}_0(\omega,k) \, , \hspace{0.5cm} \text{where} \hspace{0.5cm} \Tilde{G}_0(\omega,k) = \Tilde{G}_0(k) = \frac{1}{c_1 - c_6 k^2} .
\end{equation}
Knowing this, the evaluation of the first trace yields
\begin{align}
\begin{split}
\Tr \Big[ G_0 (\partial_x\phi)^2 \Big] &= \int d\tau \, dx \Big[ G_0(0,0) (\partial_x \phi)^2 \Big] \\ &= \int_{-\Omega}^{\Omega} \frac{d\omega}{2\pi} \int_{-\Lambda}^{\Lambda} \frac{dk}{2\pi} \frac{1}{c_1-c_6 k^2} \int d\tau \, dx \, (\partial_x\phi)^2 \\ &= \frac{\Omega}{\pi^2 \sqrt{c_1 c_6}} \artanh \Big( \sqrt{\frac{c_6}{c_1}} \Lambda\Big) \int d\tau \, dx \, (\partial_x\phi)^2 ,
\end{split}
\end{align}
where $\Omega$ is a frequency UV cutoff, while $\Lambda < \sqrt{\frac{c_1}{c_6}} = \frac{\sqrt{12}}{a}$ is a momentum UV cutoff. For the second trace, we have
\begin{align}
\begin{split}
\Tr \Big[ G_0 (\partial_x\phi)^2 G_0 (\partial_x\phi)^2 \Big] &= \int d^2X d^2X' G_0(X-X') (\partial_{x'} \phi)^2(X') G_0(X'-X) (\partial_x \phi)^2(X) \\ &= \int \frac{d^2 K}{(2\pi)^2} \widetilde{(\partial_x\phi)^2}(K) \widetilde{(\partial_x\phi)^2}(-K) \int \frac{d^2 K'}{(2\pi)^2} \Tilde{G}_0(K') \Tilde{G}_0(K'-K) \\ &\approx \int d\tau \, dx \, (\partial_x\phi)^4 \int_{-\Omega}^{\Omega} \frac{d\omega}{2\pi} \int_{-\Lambda}^{\Lambda} \frac{dk}{2\pi} \Tilde{G}_0^2(k) + \dots ,
\end{split}
\end{align}
where $X=(\tau,x)$, $X'=(\tau',x')$, $K=(\omega,k)$ (not to be confused with the Luttinger parameter) and $K'=(\omega',k')$. The last equality follows from setting $K=0$ in the loop integral, since a nonzero external momentum results in irrelevant terms. Evaluating the momentum integral yields
\begin{equation}
\Tr \Big[ G_0 (\partial_x\phi)^2 G_0 (\partial_x\phi)^2 \Big] = \frac{\Omega}{2\pi^2} \Bigg[ \frac{\Lambda}{c_1^2 - c_1 c_6 \Lambda^2} + \frac{1}{c_1 \sqrt{c_1 c_6}} \artanh\Big( \sqrt{\frac{c_6}{c_1}} \Lambda \Big) \Bigg] \int d\tau \, dx \, (\partial_x\phi)^4 .
\end{equation}

Hence, the two terms obtained by integrating out the Gaussian fluctuations of $\theta$ are
\begin{align}
\begin{split}
\frac{1}{4\pi^2} \int d\tau dx \, \dot{\phi} \hat{O}^{-1}(\phi) \dot{\phi} + \frac{1}{2} \Tr \log \hat{O}(\phi) \approx \frac{1}{4\pi^2 c_1} \int d\tau \, dx \, (\partial_{\tau}\phi)^2 - \delta c_2 \int d\tau \, dx \, (\partial_x\phi)^2 - \delta c_4 \int d\tau \, dx \, (\partial_x\phi)^4 + \dots ,
\end{split}
\end{align}
where the following two coefficients have been introduced:
\begin{align}
\begin{split}
\delta c_2 = \frac{c_8 \Omega}{2\pi^2\sqrt{c_1 c_6}} \artanh \Big( \sqrt{\frac{c_6}{c_1}} \Lambda\Big) = \frac{\sqrt{3}}{2\pi^2} a \Omega \artanh\Big( \frac{a \Lambda}{2\sqrt{3}} \Big) ,
\end{split}
\end{align}
and
\begin{align}
\begin{split}
\delta c_4 = \frac{c_8^2 \Omega}{8\pi^2} \Bigg[ \frac{\Lambda}{c_1^2 - c_1 c_6 \Lambda^2} + \frac{1}{c_1 \sqrt{c_1 c_6}} \artanh\Big( \sqrt{\frac{c_6}{c_1}} \Lambda \Big) \Bigg] = \frac{a^3 \Omega}{16 \pi^2} \Bigg[ \frac{6 a \Lambda}{12-a^2 \Lambda^2} + \sqrt{3} \artanh \Big( \frac{a \Lambda}{2\sqrt{3}} \Big) \Bigg] .
\end{split}
\end{align}

Knowing this, the action for $\phi$ becomes
\begin{align}
\begin{split}
S[\phi] = \int d\tau \, dx \, \Bigg[ \frac{1}{4\pi^2 c_1} (\partial_{\tau} \phi)^2 + (c_2-\delta c_2) (\partial_x\phi)^2 + c_3 (\partial_x^2 \phi)^2 + (c_4-\delta c_4) (\partial_x\phi)^4 + c_5 \cos(4\phi) \Bigg] + \dots .
\end{split}
\end{align}
This action describes the vicinity of the phase-separation transition, which occurs when $c_2-\delta c_2=0$ (at least at tree-level) and the order parameter $\delta \rho \sim \partial_x \phi$ condenses. From the bosonized Hamiltonian in the Luttinger liquid phase,
\begin{equation}
\hat{H} = \frac{1}{2\pi} \int dx \, \Big[ u K (\partial_x \theta)^2 + \frac{u}{K} (\partial_x \phi)^2 \Big] + \dots ,
\end{equation}
this happens when the Luttinger parameter $K$ diverges. But since the scaling dimension of the cosine operator is $\Delta_{\cos(4\phi)} = 4K$, this operator becomes very irrelevant near the critical point and it can thus be dropped. Finally, the action can be brought in a more standard form by first defining the new coefficients $\Tilde{c}_1 = 2\pi^2 c_1$, $\Tilde{c}_2 = c_2 - \delta c_2$, $\Tilde{c}_3 = 2c_3$, and $\Tilde{c}_4 = c_4-\delta c_4$, and then performing the rescalings $\tau \rightarrow (\Tilde{c}_1 \Tilde{c}_3)^{-1/2} \tau$ and $\phi \rightarrow (\Tilde{c}_1/\Tilde{c}_3)^{1/4} \phi$, which results in 
\begin{equation}
S[\phi] = \int d\tau \, dx \, \Bigg[ \frac{1}{2} (\partial_{\tau} \phi)^2 + \frac{1}{2} (\partial_x^2 \phi)^2 + \lambda_2 (\partial_x\phi)^2 + \lambda_4 (\partial_x\phi)^4 \Bigg] + \dots ,
\end{equation}
where
\begin{equation}
\lambda_2 = \frac{\tilde{c}_2}{\Tilde{c}_3} \, , \qquad \lambda_4 = \frac{\tilde{c}_4 \sqrt{\Tilde{c}_1}}{\Tilde{c}_3^{3/2}} .
\end{equation}

\subsection{Bosonization of superconducting order parameter}\label{Sec:Bosonization_SC_order_parameter}

This second section presents details of the bosonization of the lattice superconducting order parameter. Using Eq. (\ref{Eq:Exp_lattice_fermion}), we first get
\begin{align}
\begin{split}
\hat{\Delta}_r = \hat{f}_r \hat{f}_{r+1} &\approx a \Big[ e^{i k_F a} e^{i 2 k_F x} \psi_R(x) \psi_R(x+a) + e^{-i k_F a} e^{-i 2 k_F x} \psi_L(x) \psi_L(x+a) \\ &\hspace{1cm} + e^{-i k_F a} \psi_R(x) \psi_L(x+a) + e^{i k_F a} \psi_L(x) \psi_R(x+a) \Big] ,
\end{split}
\end{align}
where the first two terms correspond to the $q \sim 2k_F$ contributions, while the last two terms are the $q \sim 0$ contributions. Using Eq. (\ref{Eq:psi_eta}) on the first term yields
\begin{align}
\begin{split}
\psi_R(x) \psi_R(x+a) &= \frac{U_R^2}{2\pi \alpha} e^{-i[\phi(x)-\theta(x)]} e^{-i[\phi(x+a)-\theta(x+a)]} \\ &= \frac{U_R^2}{2\pi \alpha} e^{i\pi/2} e^{[-i(\phi(x)+\phi(x+a)) + i (\theta(x)+\theta(x+a))]} \\ &\approx  U_R^2 \frac{i}{2\pi \alpha} e^{-i 2(\phi(x)-\theta(x))} + \dots ,
\end{split}
\end{align}
where the BCH formula $e^A e^B = e^{A+B} e^{[A,B]/2}$ was used, with $A = -i[\phi(x)-\theta(x)]$ and $B = -i[\phi(x+a)-\theta(x+a)]$, which results in $[A,B] = i \pi$. Similarly, one can show that
\begin{align}
\begin{split}
\psi_L(x) \psi_L(x+a) &= \frac{U_L^2}{2\pi \alpha} e^{i[\phi(x)+\theta(x)]} e^{i[\phi(x+a)+\theta(x+a)]} \approx  U_L^2 \frac{-i}{2\pi \alpha} e^{i 2(\phi(x)+\theta(x))} + \dots ,
\end{split}
\end{align}

\begin{align}
\begin{split}
\psi_R(x) \psi_L(x+a) &= \frac{U_R U_L}{2\pi \alpha} e^{-i[\phi(x)-\theta(x)]} e^{i[\phi(x+a)+\theta(x+a)]} \approx \frac{U_R U_L}{2\pi \alpha} e^{i 2\theta(x)} + \dots ,
\end{split}
\end{align}

\begin{align}
\begin{split}
\psi_L(x) \psi_R(x+a) &= \frac{U_L U_R}{2\pi \alpha} e^{i[\phi(x)+\theta(x)]} e^{-i[\phi(x+a)-\theta(x+a)]} \approx \frac{U_L U_R}{2\pi \alpha} e^{i 2\theta(x)} + \dots .
\end{split}
\end{align}
Combining everything results in
\begin{align} \label{Eq:Exp_SC_order_parameter}
\begin{split}
\hat{f}_r \hat{f}_{r+1} &= i e^{i k_F a} e^{i 2 k_F x} \frac{U_R^2}{2\pi} e^{-i 2[\phi(x)-\theta(x)]} - i e^{-i k_F a} e^{-i 2 k_F x} \frac{U_L^2}{2\pi} e^{i 2[\phi(x)+\theta(x)]} + \frac{U_R U_L}{2\pi} \Big( e^{-i k_F a} - e^{i k_F a} \Big) e^{i 2\theta(x)} + \dots ,
\end{split}
\end{align}
where the UV cutoff $\alpha$ was set equal to the lattice spacing $a$, while the last two terms were combined using $U_L U_R = -U_R U_L$ (this accounts for the fact that $\psi_L(x) \psi_R(x+a) = - \psi_R(x+a) \psi_L(x)$). From Ref. \cite{giamarchi2003}, the scaling dimension of the vertex operator $e^{i 2\theta(x)}$ is $1/K$, while the one for $e^{i 2(\pm \phi(x)+\theta(x))}$ is $K+1/K$. Hence, near the phase transition, the first two terms in Eq. (\ref{Eq:Exp_SC_order_parameter}) can be ignored and we thus finally get
\begin{equation}
\hat{\Delta}_r \approx -\frac{i}{\pi} U_R U_L \, \sin(k_F a) e^{i 2 \theta(x)} + \dots \sim e^{i 2 \theta(x)} + \dots .
\end{equation}

\subsection{Fermion Green's function at criticality and momentum distribution}\label{Sec:Critical_fermion_G}

The Fermi ground-state momentum distribution $n(k)$ is given by the Fourier transform of the equal-time fermion single-particle Green's function:
\begin{align}\label{nk}
    n(k)&=\sum_r e^{-ikr}\langle\hat{f}_0^\dag\hat{f}_r\rangle.
\end{align}
Since the critical ground state has unbroken particle-hole symmetry, we have $\langle\hat{f}_0^\dag\hat{f}_r\rangle=(-1)^r\langle\hat{f}_0\hat{f}_r^\dag\rangle=e^{i\pi r}(\delta_{r0}-\langle\hat{f}_r^\dag\hat{f}_0\rangle)$. Using $k_F=\frac{\pi}{2}$ as well as translation symmetry $\langle\hat{f}_r^\dag\hat{f}_0\rangle=\langle\hat{f}_0^\dag\hat{f}_{-r}\rangle$, we have
\begin{align}\label{nkPHS}
    n(k_F-q)&=\sum_re^{-i(\frac{\pi}{2}-q)r}\langle\hat{f}_0^\dag\hat{f}_r\rangle\nn\\
    &=1-\sum_r e^{i(\frac{\pi}{2}+q)r}\langle\hat{f}_0^\dag\hat{f}_{-r}\rangle\nn\\
    &=1-n(k_F+q).
\end{align}
Setting $q=0$, we obtain $n(k_F)=\half$, and writing $n(k_F+q)=\half+\delta n(q)$, Eq.~(\ref{nkPHS}) further implies that $\delta n(q)$ is odd in $q$. We are interested in the long-wavelength limit $|q|\ll k_F$ of $\delta n(q)$. By rewriting Eq.~(\ref{nk}) as $n(k)=\frac{1}{L}\sum_{rr'}e^{ik(r-r')}\langle\hat{f}_r^\dag\hat{f}_{r'}\rangle$, and using the low-energy expressions Eq.~(\ref{Eq:Exp_lattice_fermion}), we find that the crossed terms $\propto\langle\psi_R^\dag\psi_L\rangle$ and $\langle\psi_L^\dag\psi_R\rangle$ are multiplied by rapidly oscillating factors $e^{\pm ik_F(x+x')}$ which are suppressed under integration over the center-of-mass coordinate $(x+x')/2$. The remaining terms can be written as
\begin{align}
    n(k_F+q)\approx\int dx\left(e^{iqx}G_R(x)+e^{i(2k_F+q)x}G_L(x)\right),
\end{align}
where we define $G_R(x)=\langle\psi_R^\dag(x)\psi_R(0)\rangle$ and $G_L(x)=\langle\psi_L^\dag(x)\psi_L(0)\rangle$. For $|q|\ll k_F$, the integral of the term $\propto G_L(x)$ converges rapidly due to the fast $e^{i2k_Fx}$ oscillations, thus any possible singular behavior in the $q\rightarrow 0$ limit is due to the first term, and we retain it only:
\begin{align}
    n(k_F+q)\approx\text{reg.}+\int_{-\infty}^\infty dx\,e^{iqx}G_R(x).
\end{align}
Using the Mandelstam formulas (\ref{Eq:psi_eta}) and the BCH formula once again, we have
\begin{align}
    G_R(x)=\frac{1}{2\pi\alpha}e^{\half[\phi(x)-\theta(x),\phi(0)-\theta(0)]}\left\langle e^{i[\phi(x)-\phi(0)-\theta(x)+\theta(0)]}\right\rangle.
\end{align}
Using the canonical commutation relations (\ref{CanonicalCommutators}), the commutator in the exponential indeed evaluates to a $c$-number,
\begin{align}
    [\phi(x)-\theta(x),\phi(0)-\theta(0)]=i\pi\sgn x,
\end{align}
and we find
\begin{align}\label{GR}
    G_R(x)=\frac{1}{2\pi\alpha}e^{\frac{i\pi}{2}\sgn x}\left\langle e^{i[\phi(x)-\phi(0)-\theta(x)+\theta(0)]}\right\rangle.
\end{align}
In a Luttinger liquid with Luttinger parameter $K$, one has $\left\langle e^{i[\phi(x)-\phi(0)-\theta(x)+\theta(0)]}\right\rangle\sim |x|^{-(K+K^{-1})/2}$~\cite{giamarchi2003}, and thus, ignoring regular terms,
\begin{align}
    \int_{-\infty}^\infty dx\,e^{iqx}G_R(x)\sim\int_0^\infty dx\,|x|^{-(K+K^{-1})/2}\sin qx
    \sim\sgn(q)\cdot|q|^{(K+K^{-1}-2)/2},
\end{align}
implying the well-known power-law singularity $n(k)-\half\sim\sgn(k-k_F)\cdot|k-k_F|^\delta$ near $k_F$, with $\delta=\half(K+K^{-1}-2)$.

In the present case, at PS criticality, the Josephson phase $\theta$ is pinned by ODLRO. Thus, we can approximate $\theta(x)$ as a constant at long distances, choosing the constant to be zero without loss of generality. Furthermore, in Ref.~\cite{andriyakhina2026}, it was argued based on a large-$N$ expansion that the correlator $\left\langle e^{i[\phi(x)-\phi(0)]}\right\rangle$ should behave as a stretched exponential:
\begin{align}
 \left\langle e^{i[\phi(x)-\phi(0)]}\right\rangle\sim|x|^{-\sigma}\exp(-C|x|^\alpha),   
\end{align}
with the stretching exponent $\alpha=\frac{2}{3}$, and where the power-law prefactor $|x|^{-\sigma}$ arises from possible logarithmic corrections to the large-$N$ saddle point. The right-moving Green's function (\ref{GR}) then becomes,
\begin{align}
    G_R(x)\sim e^{\frac{i\pi}{2}\sgn x}|x|^{-\sigma}\exp(-|x/\lambda|^\alpha),
\end{align}
where we define the length scale $\lambda\equiv C^{-1/\alpha}$, and thus
\begin{align}
    \int_{-\infty}^\infty dx\,e^{iqx}G_R(x)\sim I(q)\equiv\int_0^\infty dx\,x^{-\sigma}e^{-(x/\lambda)^\alpha}\sin qx.
\end{align}
Because the function $e^{-(x/\lambda)^\alpha}$ decays faster than any power law, all the (odd) derivatives of $n(k)$ at $k=k_F$ are given by a convergent integral and are thus finite:
\begin{align}\label{nkTaylorCoeff}
    \left.\frac{\partial^{2m+1} n(k)}{\partial k^{2m+1}}\right|_{k=k_F}\sim\left.\frac{\partial^{2m+1}I(q)}{\partial q^{2m+1}}\right|_{q=0}=(-1)^m\int_0^\infty\,dx\,x^{2m+1-\sigma}e^{-(x/\lambda)^\alpha}=\frac{(-1)^m}{\alpha}\lambda^{2m+2-\sigma}\Gamma\left(\frac{2m+2-\sigma}{\alpha}\right)<\infty.
\end{align}
Thus, unlike in a Luttinger liquid, $n(k)$ is smooth (infinitely differentiable) at $k=k_F$. However, $n(k)$ is not analytic at $k=k_F$ as can be seen from the following argument. From the derivatives in Eq.~(\ref{nkTaylorCoeff}), we construct the following Taylor series for $I(q)\sim\delta n(q)$:
\begin{align}\label{nkPowerSeries}
    I(q)&=\sum_{m=0}^\infty\frac{1}{(2m+1)!}\left.\frac{\partial^{2m+1}I(q)}{\partial q^{2m+1}}\right|_{q=0} q^{2m+1}=\sum_{m=0}^\infty a_m(q\lambda),
\end{align}
where the $q$-dependent terms in the Taylor series $a_m(q \lambda)$ are
\begin{align}
    a_m(q\lambda)=\frac{\lambda^{1-\sigma}}{\alpha}\frac{(-1)^m}{(2m+1)!}\Gamma\left(\frac{2m+2-\sigma}{\alpha}\right)(q\lambda)^{2m+1}.
\end{align}
However, we can now show that this power series diverges for any nonzero value of $q$. Consider the ratio of successive coefficients:
\begin{align}
    \c{L}_m&\equiv\left|\frac{a_{m+1}(q\lambda)}{a_m(q\lambda)}\right|=\frac{(2m+1)!}{(2m+3)!}\frac{\Gamma\left(\frac{2m+4-\sigma}{\alpha}\right)}{\Gamma\left(\frac{2m+2-\sigma}{\alpha}\right)}(q\lambda)^2=\frac{(q\lambda)^2}{(2m+3)(2m+2)}\frac{\Gamma\left(\frac{2m}{\alpha}+\frac{4-\sigma}{\alpha}\right)}{\Gamma\left(\frac{2m}{\alpha}+\frac{2-\sigma}{\alpha}\right)}.
\end{align}
The d'Alembert ratio test stipulates that the series (\ref{nkPowerSeries}) diverges if $\c{L}_m>1$ for all $m$ larger than some value. Using the asymptotic behavior of the ratio of two Gamma functions for large argument (6.1.47 in Ref.~\cite{AbramowitzStegun}),
\begin{align}
\frac{\Gamma(z+a)}{\Gamma(z+b)}\sim z^{a-b},\hspace{10mm}|z|\rightarrow\infty,
\end{align}
we find that for large $m$,
\begin{align}\label{Lm}
    \c{L}_m\sim\frac{(q\lambda)^2}{4m^2}\left(\frac{2m}{\alpha}\right)^{2/\alpha}\sim(q\lambda)^2m^{2\left(\frac{1}{\alpha}-1\right)}.
\end{align}
For any $0<\alpha<1$, including the case $\alpha=\frac{2}{3}$ predicted in Ref.~\cite{andriyakhina2026} and also observed in Fig.~\ref{fig:stretchy}, the exponent of $m$ in Eq.~(\ref{Lm}) is positive and thus $\c{L}_m$ grows with $m$ for any nonzero $q$. Thus for any nonzero $q$, an $m$ value can be found such that $\c{L}_m>1$ for all large $m$, and the series formally diverges (but functions as an asymptotic series for $|q|\lambda\ll 1$). Therefore, we conclude that the momentum distribution $n(k)$ at the PS transition features a very weak singularity at $k=k_F$: it is a nonanalytic smooth function (analogous to the free-energy singularity at a Kosterlitz-Thouless transition).

\section{Numerical methods}\label{app:numerics}

\subsection{DMRG}
\label{sec:DMRG}

Our DMRG calculations used the Julia libraries ITensorMPS.jl and ITensor.jl \cite{fishman2022}. We fixed the bond dimension to $\chi=600$ in all DMRG calculations, except when it could be eagerly reduced while maintaining the maximum truncation error of the ITensorMPS DMRG algorithm $\epsilon_{\mathrm{max}}$ below $\epsilon_{\mathrm{max}}=10^{-12}$. To enforce finding exactly one of the two PS ground states, we used particle number conservation. We found that occasionally DMRG would converge to a state that was not an eigenstate of $\hat{Q}$ when not using quantum number conservation near the critical point. We found a representative particle number curve $\langle n\rangle$ from DMRG on a system with $L=256$ without using quantum number conservation (repeating until the variance of $\hat{Q}$ was less than $1/2$) over a wide range of $g$. Interpolating this curve, we ran DMRG in particle number sectors around the curve. Starting with the best guess from the interpolated curve, we ran DMRG sweeps on an MPS with conserved particle number but with sites randomly picked to be occupied or empty to sit in a fixed $Q$ sector. We needed to know if the lowest-energy state of this particle number sector was the ground state, so we did multiple sweeps above and below the initial particle number guess in integer steps until an energy minimum was clearly located.

Enough DMRG sweeps were performed to ensure the variance in the ground-state energy,
\begin{equation}
    v_E=\frac{1}{L}\left(\langle \hat{H}^2\rangle-\langle \hat{H}\rangle^2\right),
\end{equation}
was always below $v_E \lesssim 10^{-8}$. The maximum truncation error of the ITensorMPS DMRG algorithm $\epsilon_{\mathrm{max}}$ was kept below $\epsilon_{\mathrm{max}}\lesssim 10^{-9}$ in the range $3\lesssim g \lesssim 6$, while outside of this range the maximum truncation was near machine precision $\epsilon_{\mathrm{max}} \lesssim 10^{-12}$. While our method computes the ground states of different particle number sectors, which generally converge worse in DMRG compared to the true ground state, the $Q$ sectors of interest were always near the ground state in energy and each sector still converged within $\epsilon_{\mathrm{max}}\lesssim 10^{-9}$. The difference between the ground-state energy between the second-to-last and final DMRG sweeps was $\delta E \lesssim 3\times 10^{-11}$ for all $g <6$. A select few points for $g\geq 6$ converged only to $\delta E\lesssim 5\times 10^{-8}$.

\subsection{VUMPS}
\label{sec:VUMPS}

We developed our own custom implementation of the VUMPS algorithm \cite{zauner-stauber2018} using the ITensor.jl library \cite{fishman2022} as a backbone for the tensor network operations. Our implementation did not use any quantum number conservation or multi-site unit cells, but was highly accelerated. Our results were calculated at a fixed bond dimension of $\chi=1600$, and the VUMPS convergence criterion was generally $\epsilon_{\mathrm{prec}}\lesssim 2\times 10^{-7}$. We observed that this level of convergence was sufficient for accurately calculating the expectation value $\langle\delta n_0 \rangle$ in Fig.~\ref{fig:beta} and the pairing correlator in Fig.~\ref{fig:PairingCorrelator}, and refining our results beyond this precision only increased the unphysical correlation length $\xi_\chi$ due to the finite MPS bond dimension $\chi$ towards some maximal value. Fermionic correlators, such as the Green's function in real space, required additional Jordan-Wigner string operators and generally had worse long-distance performance than the bosonic correlators, such as the pairing correlator.

\begin{figure}[t]
    \includegraphics[width=0.5\linewidth]{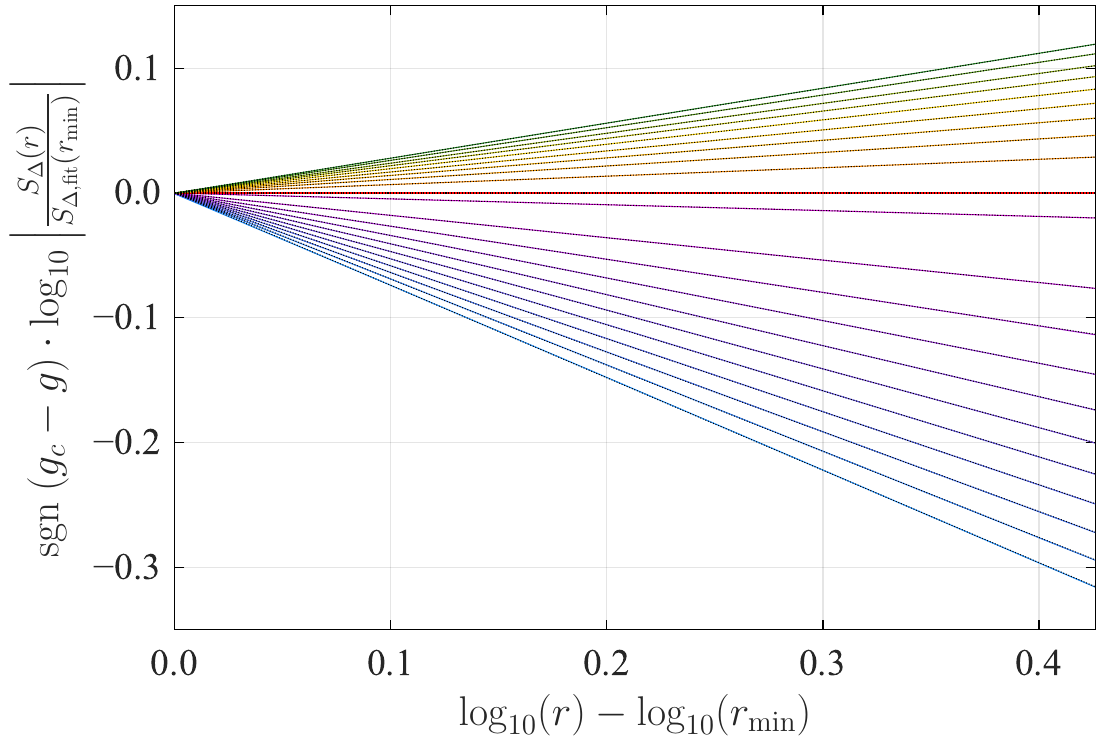}
    \caption{The power-law exponent fits used in Fig.~\ref{fig:PairingCorrelatorLL}. The sign of $g_c-g$ is used to visually separate the two phases, as they overlap otherwise. The fits are plotted as black dotted lines on top of the data, and almost lie exactly on top of the data for each $g$. The shown values of $g$ range from $g_{\min}=3.95$ (lowermost in blue) to $g_{\max}=4.14$ (uppermost in green) in steps of $\Delta g= 0.01$ with the exception of the critical point at $g_{c}=4.0514$ (flat in red), matching with the range of data in Fig.~\ref{fig:PairingCorrelatorLL}. The fits run from $r_{\min}=300$ to $r_{\max}=800$, and the correlator is scaled by the intercept of the log-log fit $S_{\Delta,\mathrm{fit}}(r_{\min})$.}
    \label{fig:PairingCorrelatorLL_Fits}
\end{figure}

\begin{figure}[t]
    \includegraphics[width=\linewidth]{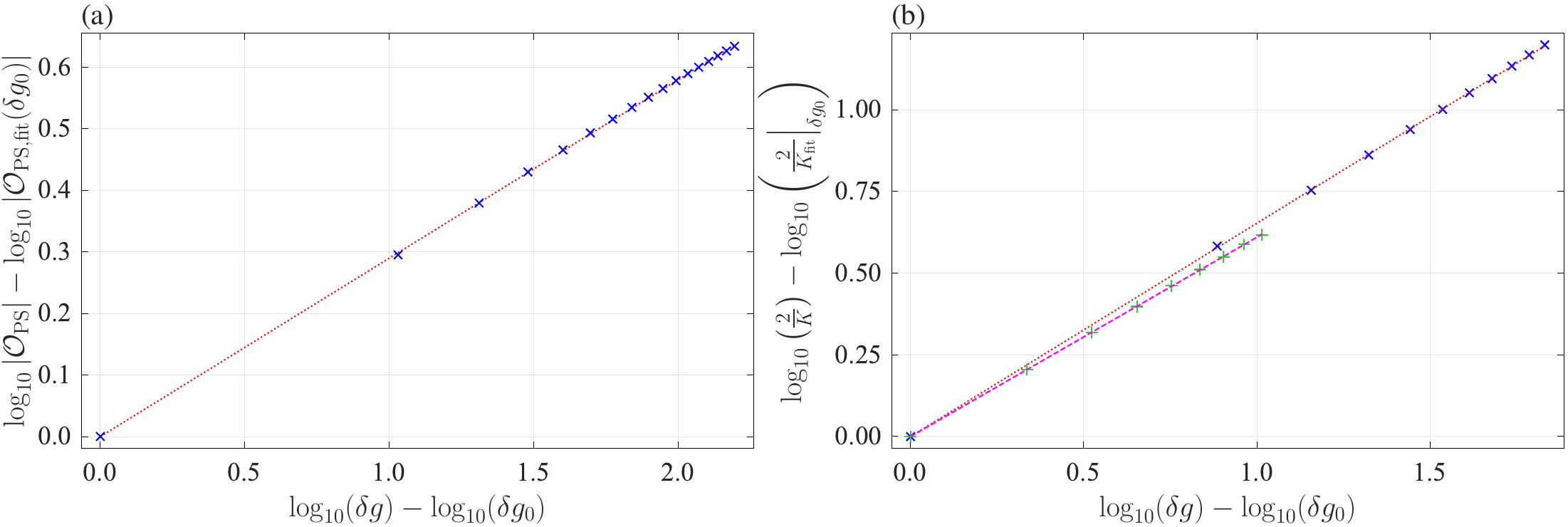}
    \caption{Log-log power law fits as a function of $\delta g\equiv |g-g_c|$, and where $\delta g_0$ is the smallest value of $\delta g$ for a given range of data. The y-axis data has been normalized to the fits' intercept. (a) The power-law fit for $\beta$ shown in Fig.~\ref{fig:beta}, fit to $g_{c,\beta}\approx4.0511$ and $\beta=0.29\pm 0.02$ with the uncertainty estimated based on the exponent's sensitivity to $g_{c,\beta}$. The blue crosses and red fit line correspond to the data $g<g_c$ in Fig.~\ref{fig:beta}. (b) The power-law fits for $\varphi$ shown in Fig.~\ref{fig:PairingCorrelatorLL}. Both fits agreed on the value $g_{c,\varphi}\approx 4.0515$, and though the slopes differ slightly, they are quite sensitive to the choice of $g_{c,\varphi}$ and can be made to agree within $\pm 0.0002$ of $g_{c,\varphi}$. We report the average exponent from the two fits as $\varphi=0.63 \pm 0.04$, with the uncertainty estimated based on the sensitivity of the fits to $g_{c,\varphi}$. The blue crosses and red dotted-line fit correspond to the data $g<g_c$ in Fig.~\ref{fig:PairingCorrelatorLL}, while the green plus markers and magenta dashed-line fit correspond to the data $g>g_c$ in Fig.~\ref{fig:PairingCorrelatorLL}.}
    \label{fig:loglog_fits}
\end{figure}

The exponents in Fig.~\ref{fig:PairingCorrelatorLL} in the main body of the paper were calculated according to the fits shown in Fig.~\ref{fig:PairingCorrelatorLL_Fits}. The logarithm of the pairing correlator $S_\Delta(r)$ in the region $r\in[300,800]$ is exceptionally linear, yielding reliable estimates for the pairing correlator power-law exponent $2/K$ despite being a fit to less than a decade of data. The power-law fits for $\beta$ and $\varphi$ are shown in Fig.~\ref{fig:loglog_fits}.

In comparison with Ref.~\cite{andriyakhina2026}, we needed to use a 16-site (instead of a four-site) moving average window to smooth out the $k_F$ oscillations in the single-particle Green's function. This moving average window was initialized to zero, causing the flat initial region in Fig.~\ref{fig:stretchy}.

\subsection{Luttinger parameter consistency check}
\label{sec:LLcheck}

\begin{figure}[t]
    \includegraphics[width=\linewidth]{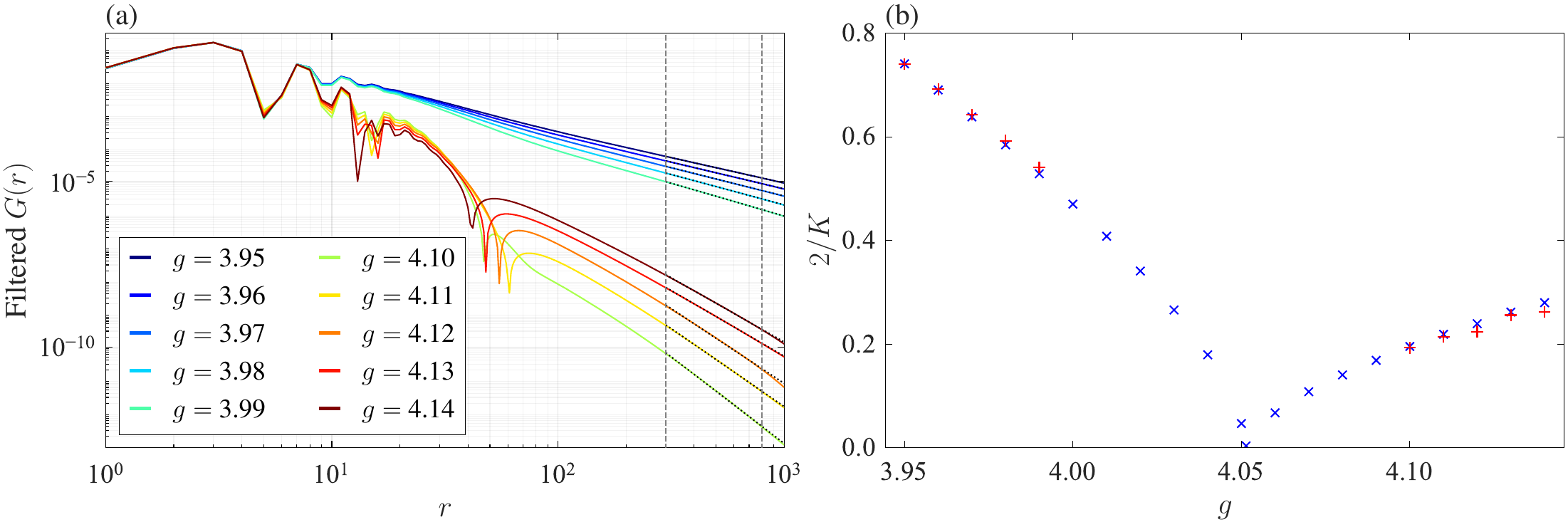}
    \caption{Validation of the Luttinger parameters calculated from the power-law fits to the pairing correlator in Fig.~\ref{fig:PairingCorrelatorLL}. (a) Fits to the filtered Green's functions for $g$ sufficiently above and below the critical point to avoid stretched-exponential behavior. The fits are plotted as dotted black lines, with the vertical dashed lines indicating the range of $r$ used for the fits. The fits are performed over the same range of $r\in[300,800]$ as in Fig.~\ref{fig:PairingCorrelatorLL_Fits}. (b) Comparison between the exponent $2/K$ calculated from the power-law fits of the pairing correlator in blue crosses, and the value $2/K$ calculated from the power-law exponents $-(K+K^{-1})/2$ fit from (a) in red plus signs. }
    \label{fig:Greens_fuction_K}
\end{figure}

The Luttinger parameter $K$, as extracted from the pairing correlator in Fig.~\ref{fig:PairingCorrelatorLL}, can also be obtained from the real-space Green's function away from the critical point. We cannot easily extract the exponent $-(K+K^{-1})/2$ from the Luttinger-liquid prediction~\cite{giamarchi2003}
\begin{equation}
    G(r)\sim \sin(k_Fr)\ |r|^{-(K+K^{-1})/2},
    \label{Eq:Greens_real_sine}
\end{equation}
near the critical point, because there the envelope of the Green's function crosses over from a power-law decay to the stretched exponential of Eq.~(\ref{StretchedExp}). However, sufficiently far from the critical point where the Green's function is well described by Eq.~(\ref{Eq:Greens_real_sine}), we can eliminate the modulation of $G(r)$ by $\sin(k_Fr)$, with $k_F=\pi n=\pi\langle\delta n_0\rangle+\frac{\pi}{2}$ given by Luttinger's theorem~\cite{luttingerFermiSurfaceSimple1960}, by multiplying $G(r)$ by $\sin{k_Fr}$ and filtering out the resulting high-frequency $\sim\sin(2k_Fr)$ oscillations. This approximately yields the power-law envelope in Eq.~(\ref{Eq:Greens_real_sine}) without any $k_F$ oscillations. From this envelope, we can perform a power-law fit for the exponent $-(K+K^{-1})/2$, solve for $K$, and compare to the $2/K$ results of Fig.~\ref{fig:PairingCorrelatorLL}. This is shown in Fig.~\ref{fig:Greens_fuction_K}, yielding reasonably consistent results between both methods.

\newpage

\twocolumngrid

\bibliography{u1ssb}

\end{document}